\documentstyle[12pt,epsf]{article}

\begin{document}

\renewcommand{\theequation}{\thesection.\arabic{equation}}
\newcommand{\reseteqnum}{\setcounter{equation}{0}}

\title{
\hfill
\parbox{4cm}{\normalsize UT-KOMABA 99-7\\hep-th/9905230}\\
\vspace{2cm}
Brane Configurations for Three-dimensional\\
Nonabelian Orbifolds
\vspace{1.5cm}}
\author{Tomomi Muto\thanks{e-mail address:
\tt muto@hep1.c.u-tokyo.ac.jp}
\vspace{0.5cm}\\
{\normalsize\em Institute of Physics}\\
{\normalsize\em University of Tokyo, Komaba, Tokyo 153, Japan}}
\date{\normalsize}
\maketitle
\vspace{1cm}

\begin{abstract}
\normalsize

We study brane configurations corresponding to D-branes
on complex three-dimensional orbifolds ${\bf C}^3/\Gamma$
with $\Gamma=\Delta(3n^2)$ and $\Delta(6n^2)$,
nonabelian finite subgroups of $SU(3)$.
We first construct a brane configuration for
${\bf C}^3/{\bf Z}_n \times {\bf Z}_n$
by using D3-branes and a web of $(p,q)$ 5-branes of type IIB string theory.
Brane configurations for the nonabelian orbifolds are
obtained by performing certain quotients
on the configuration for ${\bf C}^3/{\bf Z}_n \times {\bf Z}_n$.
Structure of the quiver diagrams of the groups
$\Delta(3n^2)$ and $\Delta(6n^2)$ can be reproduced
from the brane configurations.
We point out that the brane configuration for ${\bf C}^3/\Gamma$
can be regarded as
a physical realization of the quiver diagram of $\Gamma$.
Based on this observation,
we discuss that three-dimensional McKay correspondence
may be interpreted as T-duality.

\end{abstract}

\newpage

\section{Introduction}

D-branes at orbifold singularities have been intensively studied
for recent years (see \cite{DM}-\cite{Compean} for example).
One of the purposes is to investigate geometries of spacetime.
D-branes serve as tools to study ultrashort structure of spacetime.
So it is interesting to study geometries by using D-branes as probes
and compare them with geometries probed by point particles or
fundamental strings.
Orbifolds provide nontrivial but relatively simple
examples for such a purpose.

Another motivation to study D-branes on orbifolds
is to construct
large families of gauge theories.
Various choices of orbifolds lead to gauge theories
with various supersymmetries and dimensions.
It is well-known that gauge theories can be constructed
by using brane configurations following the work of \cite{HW}.
Investigations from different approaches
and comparison between them are useful to clarify
various aspects of gauge theories.
Investigations along this line have been made in \cite{BK,KLS,HU}
for example.

In this paper,
we study D-branes on three-dimensional nonabelian orbifolds
${\bf C}^3/\Gamma$ with $\Gamma \in SU(3)$.
For these cases, the gauge theories have $1/8$ supersymmetry
compared with type II string theory.
It is known that finite subgroups of $SU(3)$ are classified into
$ADE$-like series \cite{MBD,FFK,YY}.
We are concerned with "D"-type subgroups
$\Delta(3n^2)$ and $\Delta(6n^2)$,
where $n$ is a positive integer
and the number in braces represents the order of the group.

In studying D-brane gauge theory on ${\bf C}^3/\Gamma$,
the quiver diagram of $\Gamma$ plays an important role.
The quiver diagram of $\Gamma$ is a diagram which represents
algebraic structure of $\Gamma$.
It consists of nodes and arrows connecting the nodes.
The nodes represent irreducible representations of $\Gamma$,
while the arrows represent structure of tensor products
between a certain faithful three-dimensional representation
and the irreducible representations.
In the gauge theory, the nodes represent gauge groups and
the arrows represent matter contents.
The quiver diagrams of the groups $\Delta(3n^2)$ and $\Delta(6n^2)$
were calculated in \cite{Muto2}.
The quiver diagram of $\Delta(3n^2)$ can be considered as a ${\bf Z}_3$
quotient of that of ${\bf Z}_n \times {\bf Z}_n$,
and the quiver diagram of $\Delta(6n^2)$ is obtained by a further
${\bf Z}_2$ quotient.
In \cite{Muto2},
it was also pointed out that these quiver diagrams have similar structure
to a web of $(p,q)$5-branes of type IIB string theory \cite{Schwarz1,AHK}.
So it is natural to expect that the D-brane gauge theories can be realized
by brane configurations involving such a web of $(p,q)$ branes.
The purpose of the present paper is to construct brane configurations
corresponding to D-branes on ${\bf C}^3/\Delta(3n^2)$
and ${\bf C}^3/\Delta(6n^2)$.

In constructing brane configurations,
the brane box model \cite{HZ,HU} provides useful information.
The brane box model is a realization of D-brane gauge theory on
${\bf C}^3/{\bf Z}_n \times {\bf Z}_n$
via brane configurations.
What is remarkable on the brane box model is that
there is a correspondence between the quiver diagram of
${\bf Z}_n \times {\bf Z}_n$
and the brane configuration for
${\bf C}^3/{\bf Z}_n \times {\bf Z}_n$.
We use such a correspondence as a guideline to construct
brane configurations for ${\bf C}^3/\Gamma$.
As stated above,
the quiver diagram of $\Delta(3n^2)$ is a ${\bf Z}_3$
quotient of that of ${\bf Z}_n \times {\bf Z}_n$.
Therefore, combining with the correspondence between quiver diagrams
and brane configurations,
we expect that the brane configuration for ${\bf C}^3/\Delta(3n^2)$
is obtained from that for ${\bf C}^3/{\bf Z}_n \times {\bf Z}_n$
by a ${\bf Z}_3$ quotient.
However, we can not define a ${\bf Z}_3$ quotient
on the brane box configuration
since the ${\bf Z}_3$ is not a symmetry of the configuration.
So we can not use the brane box configuration itself
as a configuration for ${\bf C}^3/{\bf Z}_n \times {\bf Z}_n$.
Instead, we construct a brane configuration with a ${\bf Z}_3$
symmetry maintaining the correspondence with the quiver diagram
of ${\bf Z}_n \times {\bf Z}_n$.
It consists of D3-branes and a web of
$(p,q)$ 5-branes of type IIB string theory.
The brane configuration for ${\bf C}^3/\Delta(3n^2)$
is obtained from such a configuration by a ${\bf Z}_3$ quotient.
We can see that the configuration naturally reproduces the structure
of the quiver diagram of $\Delta(3n^2)$.
In other words, the brane configuration
can be considered as a physical realization of the quiver diagram.
The brane configuration for ${\bf C}^3/\Delta(6n^2)$
is obtained by a further ${\bf Z}_2$ quotient.

The quiver diagram of $\Gamma$ is also a key ingredient
in the discussion of
what is called McKay correspondence\cite{McKay}-\cite{IN}.
The McKay correspondence is a relation
between the representation theory of a finite group $\Gamma$
and the geometries of ${\bf C}^d/\Gamma$.
It was originally found for $d=2$ and $\Gamma \in SL(2,{\bf C})$.
The McKay correspondence states that the quiver diagram of $\Gamma$
coincides with a diagram which represent
intersections among exceptional divisors of
$\widetilde{{\bf C}^2/\Gamma}$.
($\widetilde{X}$ represents the minimal resolution of $X$.)
We are concerned with its generalization to three dimensions.
If $\Gamma$ is abelian,
${\bf C}^3/\Gamma$ becomes a toric variety,
and its resolution can be discussed by using toric method.
In \cite{LV}, relations between brane configurations and
toric diagrams were discussed.
Applying such arguments to the case $\Gamma={\bf Z}_n \times {\bf Z}_n$,
we can see that the brane configuration and the orbifolds
are related by T-duality, and the brane configuration is
graphically dual to the corresponding toric diagram.
Combining with an observation that
the brane configuration for ${\bf C}^3/\Gamma$
can be regarded as the quiver diagram of $\Gamma$,
we can say that the quiver diagram and the toric diagram
are related by T-duality.
Since the toric diagram represents geometric information of the orbifold
such as intersections among exceptional divisors,
it leads to an interpretation that the three-dimensional
McKay correspondence can be understood as T-duality.

The organization of this paper is as follows.
In Section 2, we review a prescription to obtain
worldvolume gauge theory of D-branes on an orbifold ${\bf C}^3/\Gamma$.
We also present quiver diagrams for $\Gamma=\Delta(3n^2)$
and $\Delta(6n^2)$ calculated in \cite{Muto2}.
In Section 3, we first review the brane box model.
Then we construct a brane configuration for
${\bf C}^3/{\bf Z}_n \times {\bf Z}_n$
which is appropriate to construct brane configuraions
for nonabelian orbifolds.
By taking its quotient,
we construct brane configurations for ${\bf C}^3/\Delta(3n^2)$
and ${\bf C}^3/\Delta(6n^2)$.
We argue that the structure of the quiver diagrams
of $\Delta(3n^2)$ and $\Delta(6n^2)$ can be explained
by the brane configurations.
In Section 4, we discuss relations between brane configurations and
toric diagrams.
Based on the discussion of \cite{LV} which relates
brane configurations and toric diagrams,
we present evidence that the three dimensional McKay correspondence
may be interpreted as T-duality.
Section 5 is devoted to discussions.

\section{Quiver diagrams for D-branes on nonabelian orbifolds}
\reseteqnum

In this section, we review the results of \cite{Muto2},
in which the quiver diagrams for D-branes on orbifolds
${\bf C}^3/\Gamma$ with $\Gamma=\Delta(3n^2)$
and $\Delta(6n^2)$ were obtained.
We start with $N$ parallel D1-branes on ${\bf C}^3$ where
$N=|\Gamma|$ is the order of $\Gamma$.
We will discuss the reason to take D1-branes in Sections 3 and 4.
The effective action of the D-branes
is given by the dimensional reduction of
ten-dimensional $U(N)$ super Yang-Mills theory to two-dimensions.
The bosonic field contents are a $U(N)$ gauge field $A$ and
four complex adjoint scalars $X^\mu$ ($\mu=1,2,3$) and $Y$.
Fermionic field contents are determined via unbroken supersymmetry,
so we do not touch upon them.
We take $\Gamma$ to act on complex three-dimensional space
corresponding to $X^\mu$.
Since $Y$ is irrelevant in the following discussions,
we set $Y=0$.
Next we project this theory onto $\Gamma$ invariant space.
The condition is expressed as
\begin{eqnarray}
R_{reg}(g) A R_{reg}(g)^{-1} &=& A, \label{eq:projectionA}\\
R_3(g)^\mu_\nu R_{reg}(g) X^\nu R_{reg}(g)^{-1} &=& X^\mu
\label{eq:projectionX}
\end{eqnarray}
where $g \in \Gamma$, $R_3$ is a three-dimensional representation
of $\Gamma$ which acts on spacetime indices $\mu$ and
$R_{reg}$ is the $N \times N$ regular representation
of $\Gamma$ which acts on Chan-Paton indices.
$R_3$ defines how $\Gamma$ acts on ${\bf C}^3$
to form the quotient singularity.
The condition (\ref{eq:projectionA})
implies that gauge fields surviving the projection
are $\Gamma$-invariant parts of $R_{reg} \otimes R_{reg}^*$.
Since the regular representation $R_{reg}$ has the following decomposition
\begin{equation}
R_{reg}=\oplus_{a=1}^r N_a R^a, \quad N_a={\rm dim}R^a
\end{equation}
where $R^a$ denotes an irreducible representation of $\Gamma$
and $r$ is the number of irreducible representations of $\Gamma$,
we obtain the following expression,
\begin{eqnarray}
R_{reg} \otimes R_{reg}^* |_{\Gamma{\rm -inv}}
&=&\oplus_{ab} N_a \bar{N_b} R^a \otimes R^{b*}
|_{\Gamma{\rm -inv}} \nonumber\\
&=&\oplus_a N_a \bar{N_a}.
\end{eqnarray}
Here we used Schur's lemma.
The equation means that the gauge symmetry of the resulting theory is
\begin{equation}
\prod_{a=1}^r U(N_a).
\end{equation}
If we consider $n$ D-branes on the orbifold,
the gauge group becomes
\begin{equation}
\prod_{a=1}^r U(n N_a).
\end{equation}

One can see the matter contents after the projection (\ref{eq:projectionX})
in a similar way.
They are $\Gamma$-invariant parts of
$R_3 \otimes R_{reg} \otimes R_{reg}^*$.
By defining the tensor product of the three-dimensional
representation $R_3$ and an irreducible representation $R^a$ as
\begin{equation}
R_3 \otimes R^a = \oplus_{b=1}^r n_{ab} R^b,
\label{eq:tensor}
\end{equation}
one obtains the following expression,
\begin{eqnarray}
R_3 \otimes R_{reg} \otimes R_{reg}^* |_{\Gamma{\rm -inv}}
&=&\oplus_{abc} n_{ab} N_a \bar{N_c} R^b \otimes R^{c*}
|_{\Gamma{\rm -inv}} \nonumber\\
&=&\oplus_{ab} n_{ab} N_a \bar{N_b}.
\end{eqnarray}
It means that $n_{ab}$ is the number of bifundamental fields which
transform as $N_a \times \bar N_b$ under $U(N_a) \times U(N_b)$.

The gauge group and the spectrum
are summarized in a quiver diagram.
A quiver diagram consists of $r$ nodes
corresponding to irreducible representations
and arrows which connect these nodes corresponding to
bifundamental matters.
Outgoing arrows represent fundamentals and ingoing arrows
represent anti-fundamentals.
$n_{ab}$ is the number of arrows
from the $a$-th node to the $b$-th node.

\subsection{$\Delta (3n^2)$ case}

The group $\Delta (3n^2)$ consists of the following elements
\begin{equation}
A_{i,j}=\left(
\begin{array}{ccc}
\omega_n^i&0&0 \\
0&\omega_n^j&0 \\
0&0&\omega_n^{-i-j}
\end{array}
\right),\quad
C_{i,j}=\left(
\begin{array}{ccc}
0&0&\omega_n^i \\
\omega_n^j&0&0 \\
0&\omega_n^{-i-j}&0
\end{array}
\right),\quad
E_{i,j}=\left(
\begin{array}{ccc}
0&\omega_n^i&0 \\
0&0&\omega_n^j \\
\omega_n^{-i-j}&0&0
\end{array}
\right)
\label{eq:ACE}
\end{equation}
where $\omega_n=e^{2\pi i/n}$ and $0 \leq i,j <n$.
The $n^2$ elements \{$A_{i,j}$\}
correspond to a three-dimensional reducible representation
of the abelian group ${\bf Z}_n \times {\bf Z}_n$.
The elements of the group $\Delta(3n^2)$
are obtained by multiplying the the following matrices
\begin{equation}
\left(
\begin{array}{ccc}
1&0&0 \\
0&1&0 \\
0&0&1
\end{array}
\right),\quad
\left(
\begin{array}{ccc}
0&0&1 \\
1&0&0 \\
0&1&0
\end{array}
\right),\quad
\left(
\begin{array}{ccc}
0&1&0 \\
0&0&1 \\
1&0&0
\end{array}
\right)
\label{eq:Z3}
\end{equation}
to the elements \{$A_{i,j}$\}.
The matrices (\ref{eq:Z3}) are the elements of the group ${\bf Z}_3$,
so the group $\Delta(3n^2)$ is isomorphic to the semidirect product
of ${\bf Z}_n \times {\bf Z}_n$ and ${\bf Z}_3$.

We first consider the cases with $n \notin 3{\bf Z}$.
The irreducible representations of $\Delta(3n^2)$ consists of
one-dimensional representations and three-dimensional representations.
There are 3 one-dimensional representations which we denote as
$R_1^\alpha$ $(\alpha=0,1,2)$.
The three-dimensional representations are labeled by
two integers $(l_1,l_2) \in {\bf Z}_n \times {\bf Z}_n$
with $(l_1,l_2) \neq (0,0)$.
Furthermore there are equivalence relations among the representations
$R_3^{(l_1,l_2)}$,
\begin{equation}
R_3^{(l_1,l_2)} \sim R_3^{(-l_1+l_2,-l_1)}
\sim R_3^{(-l_2,l_1-l_2)}.
\label{eq:relation}
\end{equation}
Thus the three-dimensional representations are labeled by the
lattice points $({\bf Z}_n \times {\bf Z}_n-(0,0))/{\bf Z}_3$
where the action of ${\bf Z}_3$ is defined by
the relation (\ref{eq:relation}).
By counting the number of the lattice points in
$({\bf Z}_n \times {\bf Z}_n-(0,0))/{\bf Z}_3$
one can see that there are $(n^2-1)/3$ three-dimensional
irreducible representations.
Thus the gauge group of the D-brane worldvolume theory
on ${\bf C}^3/\Delta(3n^2)$ is
$U(1)^3 \times U(3)^{(n^2-1)/3}$.

Now we choose the three-dimensional representation acting on the
spacetime indices to be $R_3^{(m_1,m_2)}$.
By calculating tensor products between $R_3^{(m_1,m_2)}$
and irreducible representations of $\Delta(3n^2)$,
we can depict the quiver diagram of $\Delta(3n^2)$ in a closed form.
An example will be given in Section 3.
However, it becomes very complicated as $n$ increases,
so it is not useful to analyze structure of the gauge theories.
Therefore we would like to depict the quiver diagram of $\Delta(3n^2)$
in a form such that its structure becomes simple.
To this end, it is useful to define $R_3^{(0,0)}$ as
\begin{equation}
R_3^{(0,0)} \equiv R_1^0 \oplus
R_1^1 \oplus R_1^2.
\label{eq:definition}
\end{equation}
Then the tensor products of $R_3^{(m_1,m_2)}$ and
the irreducible representations are given by a compact form,
\begin{equation}
R^{(m_1,m_2)}_3 \otimes R^{(l_1,l_2)}_3
= R^{(l_1+m_1,l_2+m_2)}_3
\oplus R^{(l_1-m_2,l_2+m_1-m_2)}_3
\oplus R^{(l_1-m_1+m_2,l_2-m_1)}_3.
\label{eq:a2}
\end{equation}
Although $R_3^{(0,0)}$ is not an irreducible representation,
the expression (\ref{eq:a2}) is so simple
that we tentatively forget the structure (\ref{eq:definition})
and treat it as if it were an irreducible representation.

\begin{figure}[htdp]
\begin{center}
\begin{minipage}{60mm}
\begin{center}
\leavevmode
\epsfxsize=50mm
\epsfbox{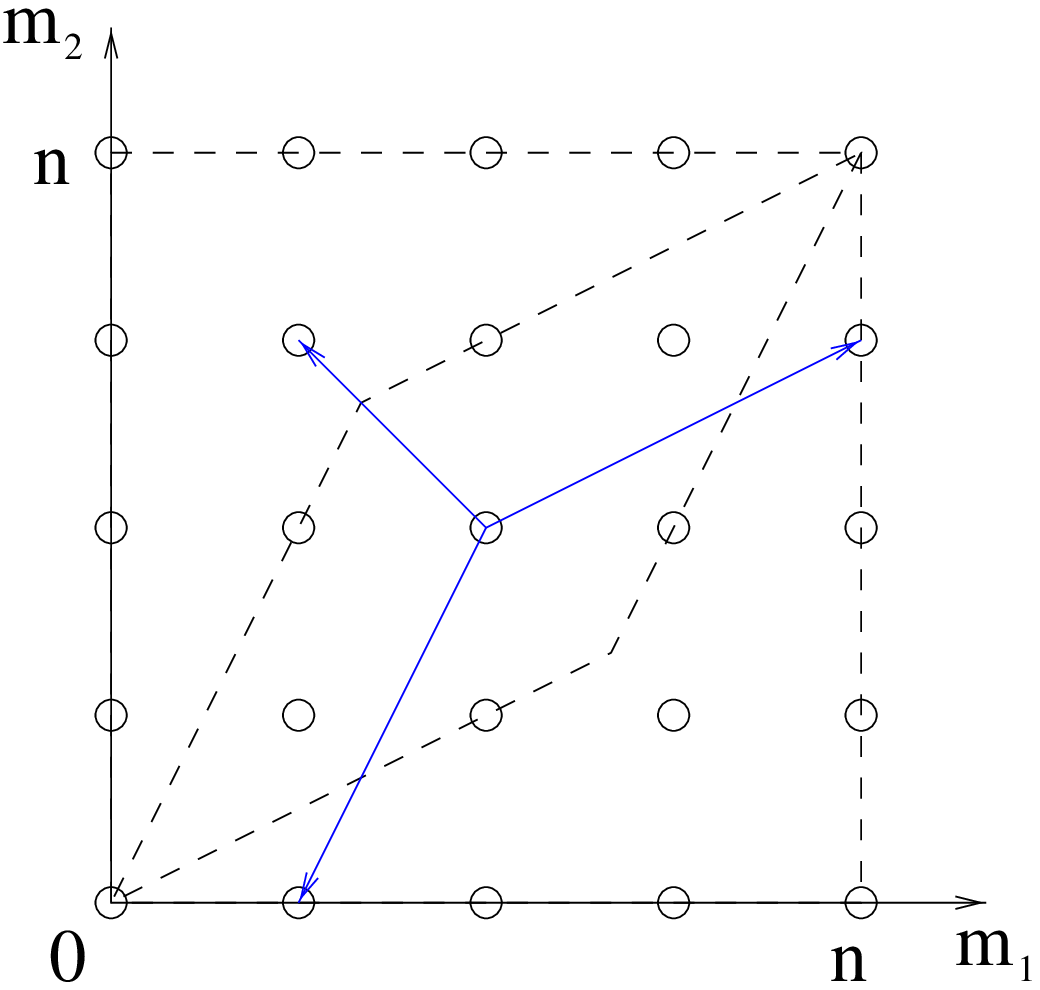}
\caption{The quiver diagram for ${\bf C}^3/\Delta(3n^2)$
when $n/3$ is not an integer.}
\label{fig:quiver1}
\end{center}
\end{minipage}
\hspace{20mm}
\begin{minipage}{60mm}
\begin{center}
\leavevmode
\epsfxsize=50mm
\epsfbox{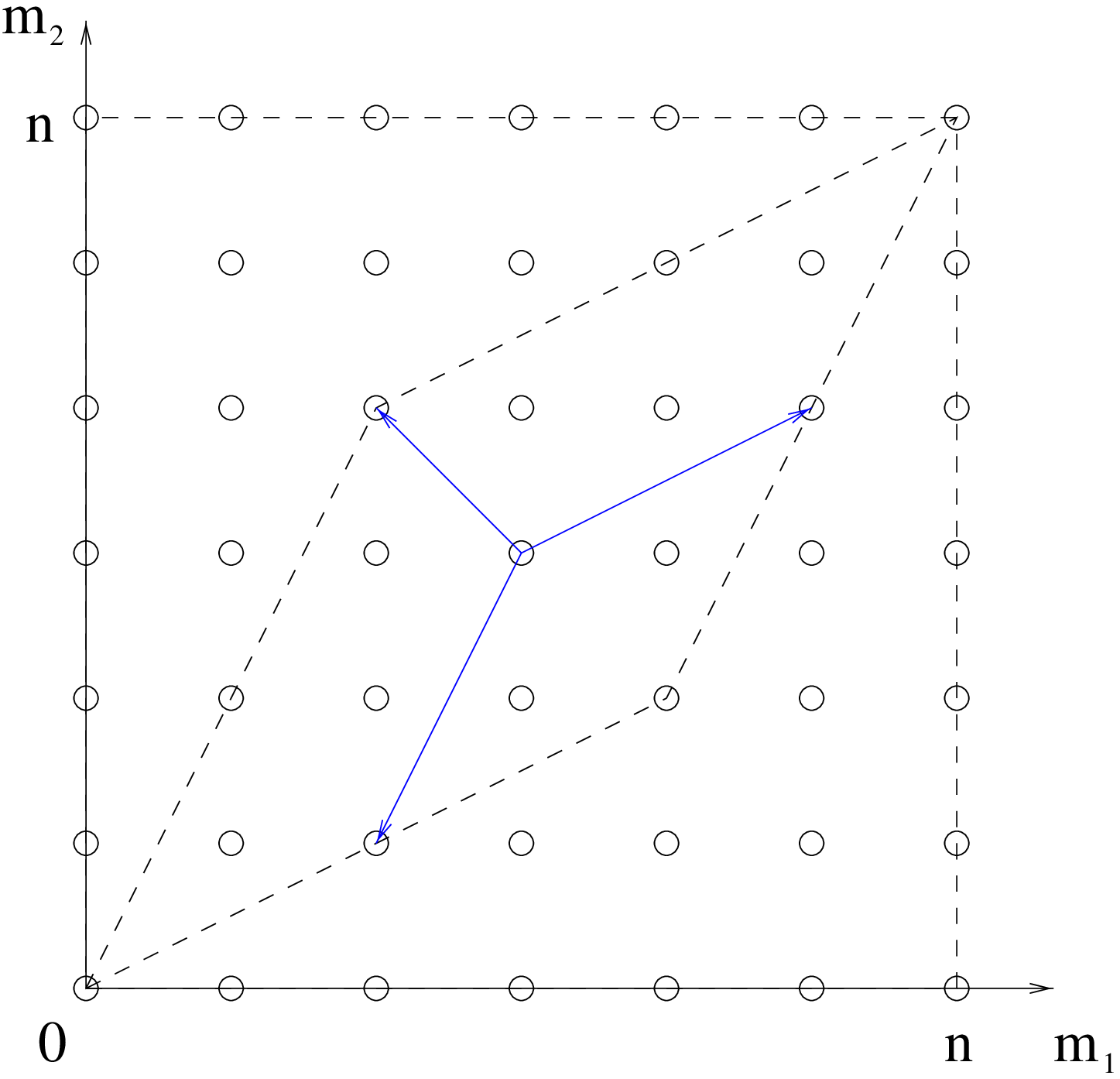}
\caption{The quiver diagram for ${\bf C}^3/\Delta(3n^2)$
when $n/3$ is an integer.}
\label{fig:quiver2}
\end{center}
\end{minipage}
\end{center}
\end{figure}

As stated above, a quiver diagram consists of nodes
associated with the irreducible representations
and arrows associated with the structure of the tensor products.
In the present case, the irreducible representations
are labeled by two integers modulo $n$,
so we put the nodes on the lattice points of ${\bf Z}_n \times {\bf Z}_n$.
The equation (\ref{eq:a2}) indicates that there are three arrows
which start from each node $(l_1,l_2)$.
The end points are $(l_1+m_1,l_2+m_2)$,
$(l_1-m_2,l_2+m_1-m_2)$ and $(l_1-m_1+m_2,l_2-m_1)$.
The quiver diagram is obtained by putting such arrows
to each node in ${\bf Z}_n \times {\bf Z}_n$
and identifying the nodes according to the ${\bf Z}_3$
equivalence relations (\ref{eq:relation}).
Thus the quiver diagram of $\Delta(3n^2)$ is basically
a ${\bf Z}_3$ quotient of that of ${\bf Z}_n \times {\bf Z}_n$.
It is depicted in Figure \ref{fig:quiver1}.
Only one set of three arrows is depicted for simplicity
although such a set of arrows start from every node.
Due to the ${\bf Z}_3$ identification,
the fundamental region is restricted to the parallelogram
surrounded by dashed lines.
To be precise, however,
we must take the equation (\ref{eq:definition}) into consideration.
Except for the fixed points,
${\bf Z}_3$ quotient means that three
one-dimensional nodes related by ${\bf Z}_3$ must be identified.
Such nodes become three-dimensional after the ${\bf Z}_3$ quotient.
For the fixed points such as the node at the origin,
the ${\bf Z}_3$ quotient
must be understood as a splitting into three one-dimensional nodes
as indicated by the equation (\ref{eq:definition}).
The rule of the ${\bf Z}_3$ quotient on arrows is defined
according to the rule on nodes
since each arrow is determined
by specifying the starting node and the ending node.

For the cases with $n \in 3{\bf Z}$,
there are 9 one-dimensional representations
which we denote as $R_1^\alpha$ $(\alpha=0,1,\dots, 8)$
and three-dimensional representations labeled by
$(l_1,l_2) \in {\bf Z}_n \times {\bf Z}_n-F$
where $F=\{(0,0), (n/3,2n/3),(2n/3,n/3)\}$.
As in the $n \notin 3{\bf Z}$ cases, there are equivalence
relations (\ref{eq:relation}).
By counting the number of the lattice points in
$({\bf Z}_n \times {\bf Z}_n-F)/{\bf Z}_3$,
one can see that there are $(n^2-3)/3$ three-dimensional
irreducible representations when $n/3$ is an integer.
So the gauge group of the D-brane theory is
$U(1)^9 \times U(3)^{(n^2-3)/3}$.

By defining
\begin{eqnarray}
&&R_3^{(0,0)} \equiv R_1^0 \oplus
R_1^1 \oplus R_1^2,\label{eq:definition-1}\\
&&R_3^{(n/3,2n/3)} \equiv R_1^3 \oplus
R_1^4 \oplus R_1^5,\label{eq:definition-2}\\
&&R_3^{(2n/3,n/3)} \equiv R_1^6 \oplus
R_1^7 \oplus R_1^8,\label{eq:definition-3}
\end{eqnarray}
tensor products of $R_3^{(m_1,m_2)}$ and
the irreducible representations are given as follows,
\begin{equation}
R^{(m_1,m_2)}_3 \otimes R^{(l_1,l_2)}_3
= R^{(l_1+m_1,l_2+m_2)}_3
\oplus R^{(l_1-m_2,l_2+m_1-m_2)}_3
\oplus R^{(l_1-m_1+m_2,l_2-m_1)}_3.
\end{equation}
The quiver diagram is shown in Figure \ref{fig:quiver2}.
It is essentially the same as in the
$n \notin 3{\bf Z}$ cases except that the nodes
$(n/3,2n/3)$ and $(2n/3,n/3)$ lie at fixed points
of the ${\bf Z}_3$ action.

\subsection{$\Delta (6n^2)$ case}

We now turn to the group $\Delta (6n^2)$.
The group $\Delta(6n^2)$ with $n$ a positive integer
consists of \{$A_{i,j}$, $C_{i,j}$, $E_{i,j}$\} in (\ref{eq:ACE})
and the following matrices.
\begin{equation}
B_{i,j}=\left(
\begin{array}{ccc}
\omega_n^i&0&0 \\
0&0&\omega_n^j \\
0&\omega_n^{\frac{n}{2}-i-j}&0
\end{array}
\right),\;
D_{i,j}=\left(
\begin{array}{ccc}
0&\omega_n^i&0 \\
\omega_n^j&0&0 \\
0&0&\omega_n^{\frac{n}{2}-i-j}
\end{array}
\right),\;
F_{i,j}=\left(
\begin{array}{ccc}
0&0&\omega_n^i \\
0&\omega_n^j&0 \\
\omega_n^{\frac{n}{2}-i-j}&0&0
\end{array}
\right). \nonumber
\label{eq:BDF}
\end{equation}
As pointed out in \cite{FFK},
even if one choose $n$ to be an odd integer,
the matrices generate elements of $\Delta(6(2n)^2)$
by multiplication.
So we restrict the value of $n$ to even integers.
The elements of the group $\Delta(6n^2)$
are obtained by multiplying the matrices (\ref{eq:Z3}) and
\begin{equation}
\left(
\begin{array}{ccc}
-1&0&0 \\
0&0&-1 \\
0&-1&0
\end{array}
\right),\quad
\left(
\begin{array}{ccc}
0&-1&0 \\
-1&0&0 \\
0&0&-1
\end{array}
\right),\quad
\left(
\begin{array}{ccc}
0&0&-1 \\
0&-1&0 \\
-1&0&0
\end{array}
\right)
\label{eq:S3}
\end{equation}
to the elements \{$A_{i,j}$\} in (\ref{eq:ACE}).
Note that the six matrices (\ref{eq:Z3}) and (\ref{eq:S3})
are the elements of the symmetric group ${\bf S}_3$
of order six.
Thus the group $\Delta(6n^2)$ is isomorphic to the semidirect product
of ${\bf Z}_n \times {\bf Z}_n$ and ${\bf S}_3$.

When $n/3$ is not an integer,
the irreducible representations consist of
2 one-dimensional representations,
1 two-dimensional representation,
$2(n-1)$ three-dimensional representations
and $(n^2-3n+2)/6$ six-dimensional representations.
Thus the gauge group is
$U(1)^2 \times U(2) \times U(3)^{2(n-1)}
\times U(6)^{(n^2-3n+2)/6}$.
We denote the one-dimensional representations as
$R_1^t$ $(t \in {\bf Z}_2)$,
the two-dimensional representation as $R_2$
and the three-dimensional representations as
$R_3^{(l,l)t}$, where $l=1,2,\dots, n-1$
and $t$ takes values in ${\bf Z}_2$.
The six-dimensional representations are labeled by
$(l_1,l_2) \in {\bf Z}_n \times {\bf Z}_n-F'$
where $F'=\{(0,0), (l,l), (l,0), (0,l)\}$.
As in the $\Delta (3n^2)$ cases,
there are equivalence relations
among the representations $R^{(l_1,l_2)}_6$,
\begin{equation}
R_6^{(l_1,l_2)} \sim R_6^{(-l_1+l_2,-l_1)}
\sim R_6^{(-l_2,l_1-l_2)},
\label{eq:relation6-1}
\end{equation}
\begin{equation}
R_6^{(l_1,l_2)} \sim R_6^{(l_2,l_1)}.
\label{eq:relation6-2}
\end{equation}
Hence the six-dimensional representations are labeled by the lattice points
$({\bf Z}_n \times {\bf Z}_n-F')/{\bf S}_3$.
Here ${\bf S}_3$ is the semidirect product of groups
${\bf Z}_3$ and ${\bf Z}_2$ whose action are defined by the
relations (\ref{eq:relation6-1}) and (\ref{eq:relation6-2}) respectively.
$(n^2-3n+2)/6$ is the number of the lattice points
in $({\bf Z}_n \times {\bf Z}_n-F')/{\bf S}_3$
when $n/3$ is not an integer.

\begin{figure}[htdp]
\begin{center}
\begin{minipage}{60mm}
\begin{center}
\leavevmode
\epsfxsize=50mm
\epsfbox{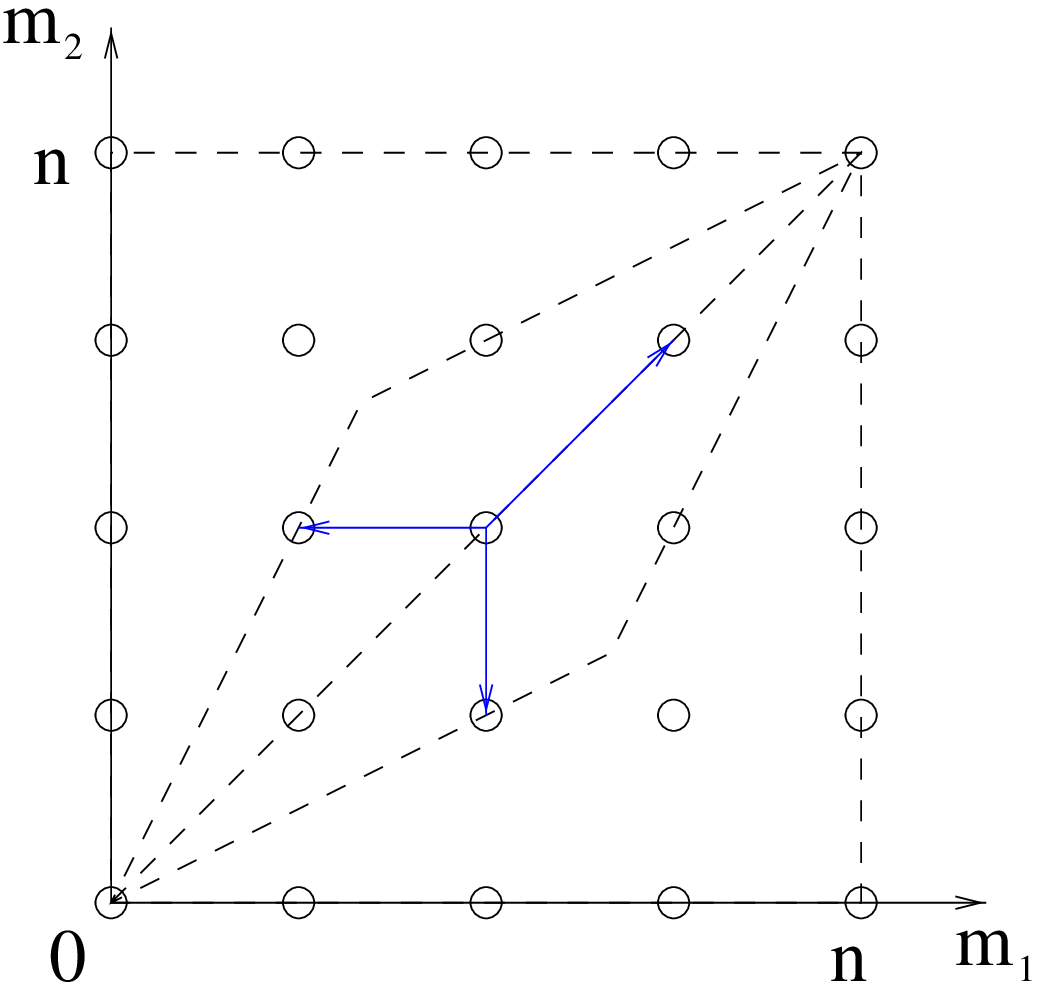}
\caption{The quiver diagram for ${\bf C}^3/\Delta(6n^2)$
when $n/3$ is not an integer.}
\label{fig:quiver3}
\end{center}
\end{minipage}
\hspace{20mm}
\begin{minipage}{60mm}
\begin{center}
\leavevmode
\epsfxsize=50mm
\epsfbox{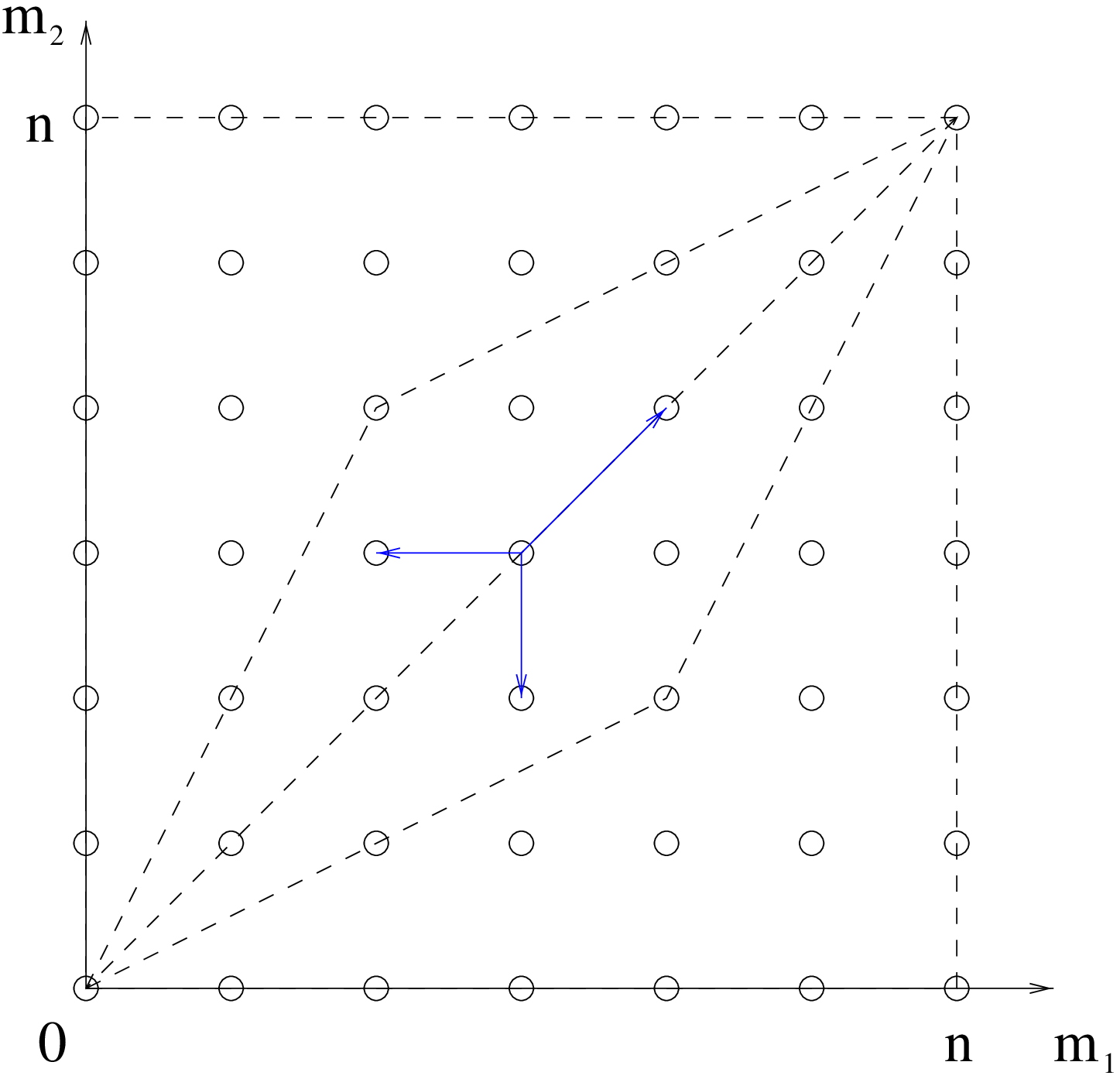}
\caption{The quiver diagram for ${\bf C}^3/\Delta(6n^2)$
when $n/3$ is an integer.}
\label{fig:quiver4}
\end{center}
\end{minipage}
\end{center}
\end{figure}

By defining
\begin{eqnarray}
&&R_6^{(0,0)} \equiv R_1^0 \oplus
R_1^1 \oplus 2 R_2,\\
&&R_6^{(l,l)}
\equiv R_3^{(l,l)0} \oplus R_3^{(l,l)1},\quad (l \neq 0)
\end{eqnarray}
tensor products of $R^{(m,m)t}_3$ and
the irreducible representations are given by,
\begin{equation}
R^{(m,m)t}_3 \otimes R^{(l_1,l_2)}_6
= R^{(l_1+m,l_2+m)}_6 \oplus R^{(l_1,l_2-m)}_6
\oplus R^{(l_1-m,l_2)}_6.
\end{equation}
The quiver diagram is depicted in Figure \ref{fig:quiver3}.
Due to the additional ${\bf Z}_2$ identification of (\ref{eq:relation6-2}),
the fundamental region is restricted to, say, the lower triangle
surrounded by dashed lines in Figure \ref{fig:quiver3}.
We note that the nodes at $(l,l)$ correspond to
fixed points of the ${\bf Z}_2$ action defined by (\ref{eq:relation6-2}).
In addition, the node at the origin corresponds to
a fixed point of the ${\bf Z}_3$ action defined by (\ref{eq:relation6-1}).

When $n/3$ is an integer,
the irreducible representations consist of
2 one-dimensional representations,
4 two-dimensional representations,
$2(n-1)$ three-dimensional representations
and $(n^2-3n)/6$ six-dimensional representations.
So the gauge group is
$U(1)^2 \times U(2)^4 \times U(3)^{2(n-1)}
\times U(6)^{(n^2-3n)/6}$.
We denote the two-dimensional representations as $R_2^\alpha$
$(\alpha=0,\dots,3)$
and use the same notations as the $n \notin 3 {\bf Z}$ case
for other representations.
As in the $n \notin 3{\bf Z}$ case, there are equivalence
relations (\ref{eq:relation6-1}) and (\ref{eq:relation6-2}),
so the six-dimensional representations are labeled by the lattice points
$({\bf Z}_n \times {\bf Z}_n-F'')/{\bf S}_3$ where
$F''=F \cup F'$.
$(n^2-3n)/6$ is the number of the lattice points
in $({\bf Z}_n \times {\bf Z}_n-F'')/{\bf S}_3$
when $n/3$ is an integer.

By defining
\begin{eqnarray}
&&R_6^{(0,0)} \equiv R_1^0 \oplus
R_1^1 \oplus 2 R_2^0,\\
&&R_6^{(l,l)}
\equiv R_3^{(l,l)0} \oplus R_3^{(l,l)1},\quad (l \neq 0)\\
&&R_6^{(2n/3,n/3)}
\equiv R_2^1 \oplus R_2^2 \oplus R_2^3,
\end{eqnarray}
tensor products of $R_3^{(m,m)t}$ and
the irreducible representations are given as follows,
\begin{equation}
R^{(m,m)t}_3 \otimes R^{(l_1,l_2)}_6
= R^{(l_1+m,l_2+m)}_6 \oplus R^{(l_1,l_2-m)}_6
\oplus R^{(l_1-m,l_2)}_6.
\end{equation}
The quiver diagram is depicted in Figure \ref{fig:quiver4}.
It is essentially the same as in the
$n \notin 3{\bf Z}$ cases except that the node
$(2n/3,n/3)$ lies at a fixed point of the ${\bf Z}_3$ action.

\section{Brane configurations for nonabelian orbifolds}
\reseteqnum

In the last section, we have presented the quiver diagrams for
D-branes on orbifolds ${\bf C}^3/\Gamma$ with $\Gamma=\Delta(3n^2)$
and $\Delta(6n^2)$.
Unexpectedly the quiver diagrams have a resemblance to
the structure of string junctions \cite{Schwarz1}
or webs of $(p,q)$ 5-branes \cite{AHK}.
Three $(p,q)$ strings of type IIB theory are permitted to form
a prong if the $(p,q)$ charge is conserved,
\begin{equation}
\sum_{i=1}^3 p_i= \sum_{i=1}^3 q_i=0.
\end{equation}
In order to have a quarter supersymmetry,
$(p,q)$ strings is constrained to have a slope $p+\tau q$
on a plane where $\tau=\frac{i}{g_s}+\frac{\chi}{2\pi}$,
$g_s$ is the coupling constant and $\chi$ is the axion
of type IIB string theory.
Similar conditions must be satisfied
to form a prong of $(p,q)$ 5-branes.
The arrows in the quiver diagrams representing matter contents
have the same structure.
It is likely that the gauge theory represented by the quiver
diagram have something to do with string junctions or a web of
$(p,q)$ 5-branes.
In this section, we consider realizations of the gauge theories
of D-branes on the nonabelian orbifolds by using such
brane configurations.

To investigate the brane configurations for nonabelian
orbifolds ${\bf C}^3/\Delta$,
we first review the brane box model,
which is a brane configuration corresponding to
D-branes on an abelian orbifold ${\bf C}^3/{\bf Z}_n \times {\bf Z}_n$
\cite{HZ,HU}.
As noted in the last section,
the finite groups $\Delta(3n^2)$ and $\Delta(6n^2)$
are closely related to ${\bf Z}_n \times {\bf Z}_n$.
We study the brane configurations
for ${\bf C}^3/\Delta$ based on the brane box model
by using relations
between the groups $\Delta$ and ${\bf Z}_n \times {\bf Z}_n$.

\begin{figure}[htdp]
\begin{center}
\begin{minipage}{60mm}
\begin{center}
\leavevmode
\epsfxsize=50mm
\epsfbox{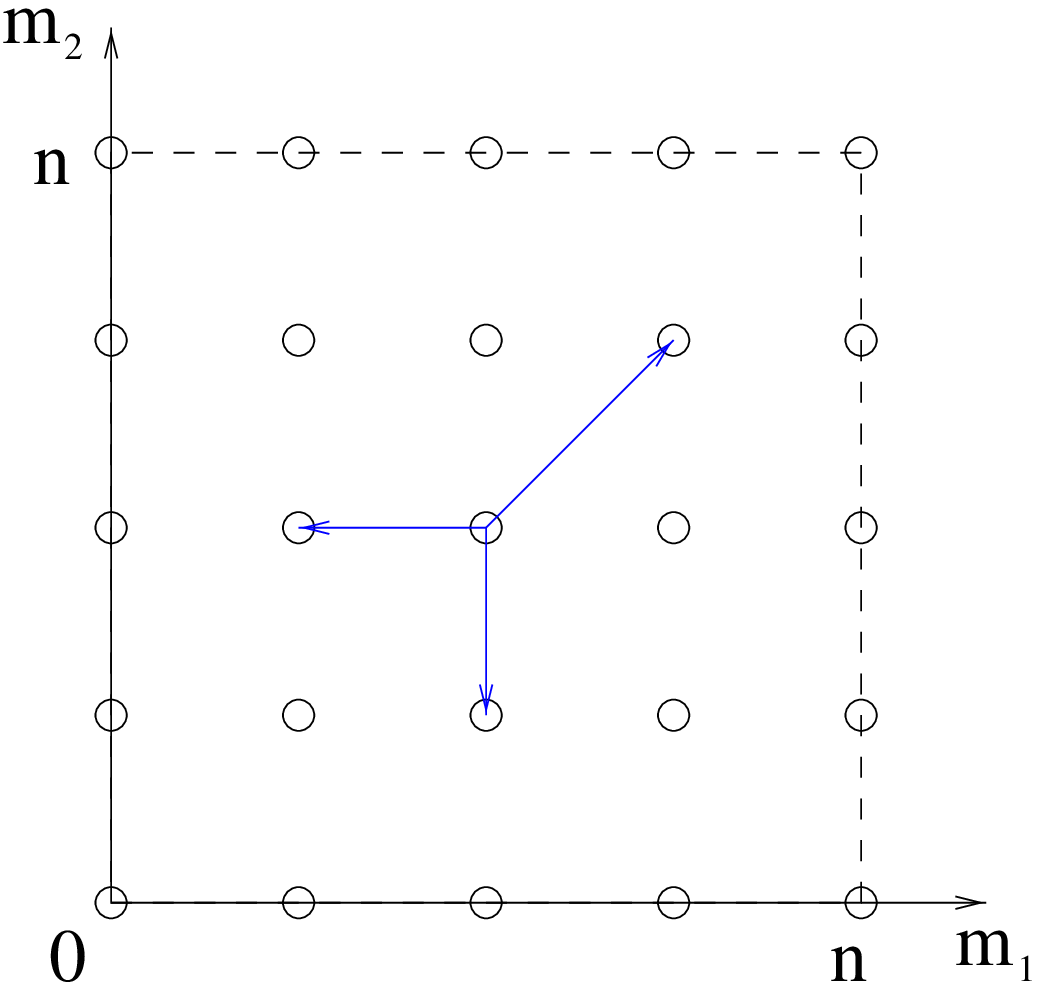}
\caption{The quiver diagram for ${\bf C}^3/{\bf Z}_n \times {\bf Z}_n$.}
\label{fig:quiver5}
\end{center}
\end{minipage}
\hspace{20mm}
\begin{minipage}{60mm}
\begin{center}
\leavevmode
\epsfxsize=50mm
\epsfbox{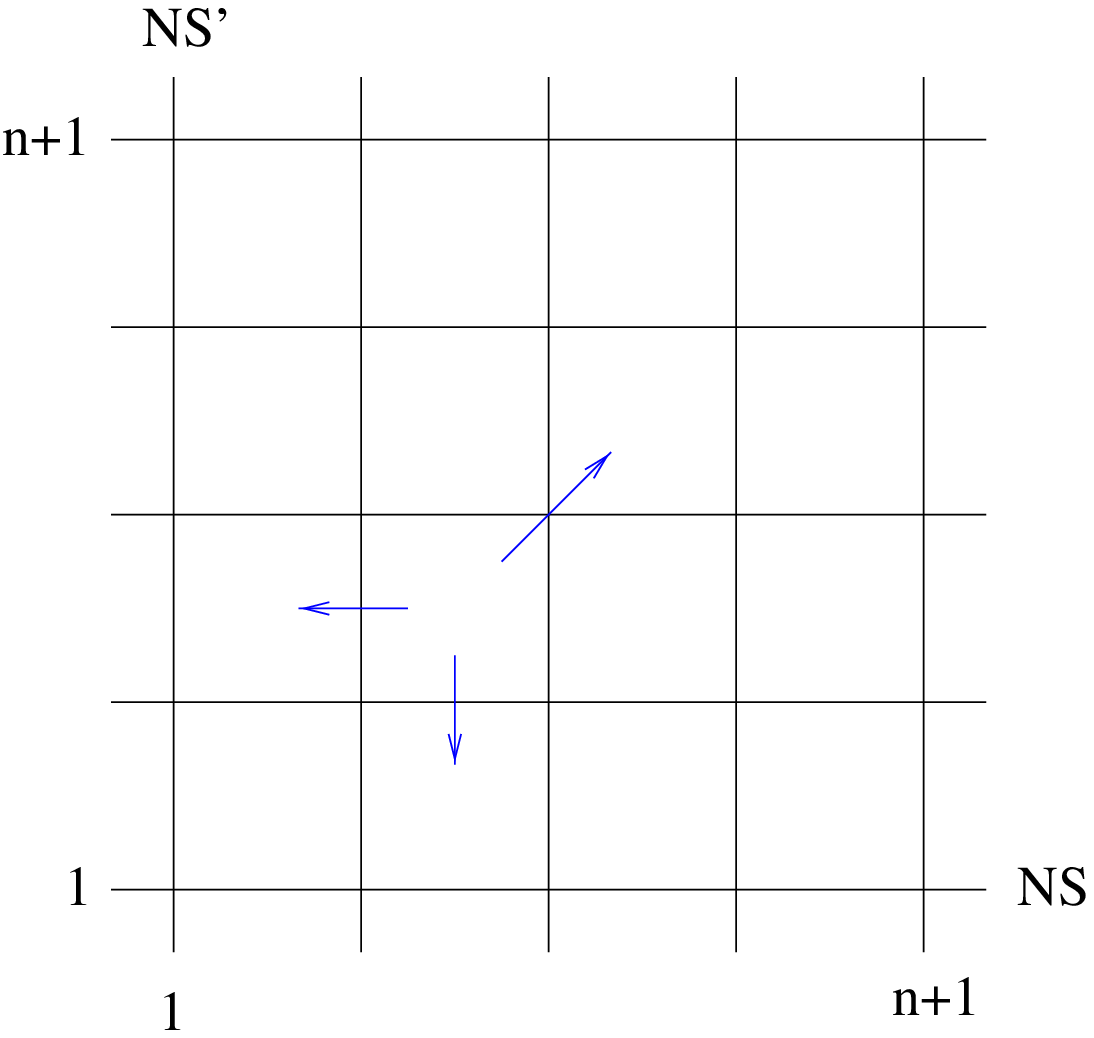}
\caption{The brane box configuration.}
\label{fig:branebox}
\end{center}
\end{minipage}
\end{center}
\end{figure}

Irreducible representations of the finite group
${\bf Z}_n \times {\bf Z}_n$ consist of $n^2$
one-dimensional representations $R_1^{(l_1,l_2)}$
with $(l_1,l_2) \in {\bf Z}_n \times {\bf Z}_n$.
We choose the three-dimensional representation
which defines the geometry of the orbifold to be
\begin{equation}
R_3 = R_1^{(1,1)}
\oplus R_1^{(-1,0)}
\oplus R_1^{(0,-1)},
\end{equation}
then the decomposition of the product $R_3 \otimes R_1^{(l_1,l_2)}$
becomes
\begin{equation}
R_3 \otimes R^{(l_1,l_2)}_1
= R^{(l_1+1,l_2+1)}_1
\oplus R^{(l_1-1,l_2)}_1
\oplus R^{(l_1,l_2-1)}_1. \label{eq:a3}
\end{equation}
The quiver diagram is given in Figure \ref{fig:quiver5}.
The point we would like to emphasize is that
the quiver diagram for the orbifold
${\bf C}^3/\Delta(3n^2)$ (${\bf C}^3/\Delta(6n^2)$) is essentially
that for the orbifold ${\bf C}^3/{\bf Z}_n \times {\bf Z}_n$ with
the ${\bf Z}_3$ equivalence relation (\ref{eq:relation})
(the $S_3$ equivalence relation
(\ref{eq:relation6-1}) and (\ref{eq:relation6-2})).

The brane box is a model that provides the same gauge theory
as the D-brane worldvolume theory on ${\bf C}^3/{\bf Z}_n
\times {\bf Z}_n$.
It consists of the following set of branes;
NS5-branes along 012345 directions,
NS'5-branes along 012367 directions and
D5-branes along 012346 directions. 
The brane box configuration is shown in Figure \ref{fig:branebox}.
It represents the brane configuration on the 46-plane.
NS and NS' 5-branes are indicated by horizontal and vertical lines
respectively.
D5-branes lie on each box bounded by NS and NS' 5-branes.
In the brane box model,
the first NS5(NS'5)-brane and the (n+1)-th NS5(NS'5)-branes
must be identified.
The interest focuses on the gauge theory
on the world-volume of the D5-branes.
Being finite in 46 directions,
the D5-branes are macroscopically (3+1)-dimensional.
Each box provides a gauge group $U(N_a)$ where $N_a$ denotes the
number of D-branes on the $a$-th box.
Open strings connecting D-branes on neighboring boxes
provide matters.
In the absence of NS5 and NS'5-branes, two possible orientations
are allowed for the strings.
The orientation of NS5 and NS'5-branes induces a particular
orientation for the strings.
One possible choice of orientations of strings is shown in Figure
\ref{fig:branebox} \cite{HZ}.
Only one set of three open strings is drawn
although such a set of strings stretch from every box.
The oriented open strings stretching from D-branes on the $a$-th box
to D-branes on the $b$-th box provide bifundamental matters
$N_a \times \bar{N_b}$.
Thus the matter contents of the brane box model
are just what determined by the quiver diagram
in Figure \ref{fig:quiver5}.
We make a list of correspondence among representation theory,
gauge theory, quiver diagram and brane box model
in Table \ref{table:correspondence}.
\begin{table}[htdp]
\caption{The correspondence among representation theory,
gauge theory, quiver diagram and brane box model.}
\label{table:correspondence}
\begin{center}
\begin{tabular}{|c|c|c|}
\hline
representation theory&$N_a$-dim irreducible repr.
&$R_3 \otimes R^a=\oplus n_{ab}R^b$ \\ \hline
gauge theory&$U(N_a)$&matters $N_a \times \bar{N_b}$\\ \hline
quiver diagram&node $a$&arrows from $a$ to $b$\\ \hline
brane box model&box with $N_a$ D-branes
&oriented open strings\\ \hline
\end{tabular}
\end{center}
\end{table}

\begin{figure}[htdp]
\begin{center}
\begin{minipage}{60mm}
\begin{center}
\leavevmode
\epsfxsize=50mm
\epsfbox{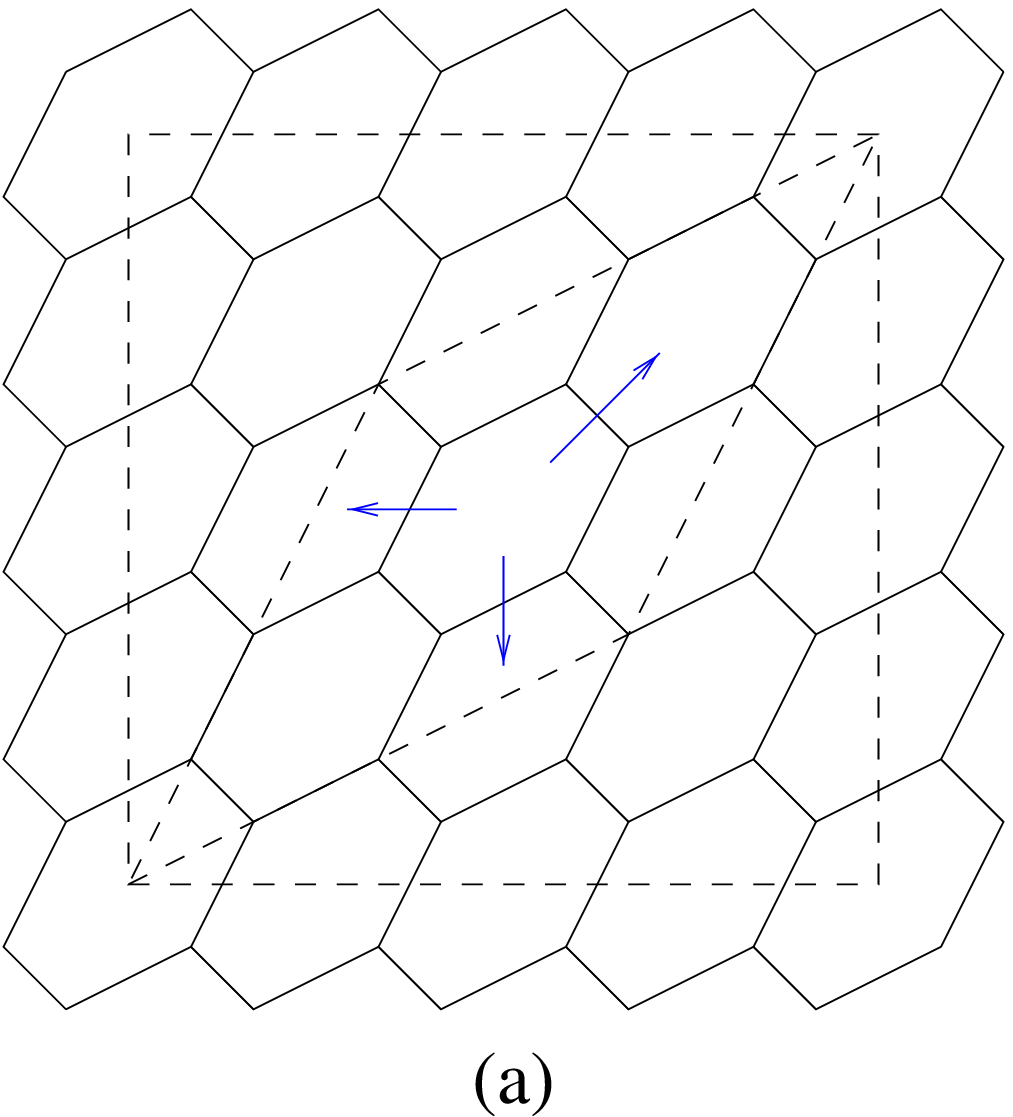}
\end{center}
\end{minipage}
\hspace{20mm}
\begin{minipage}{60mm}
\begin{center}
\leavevmode
\epsfxsize=50mm
\epsfbox{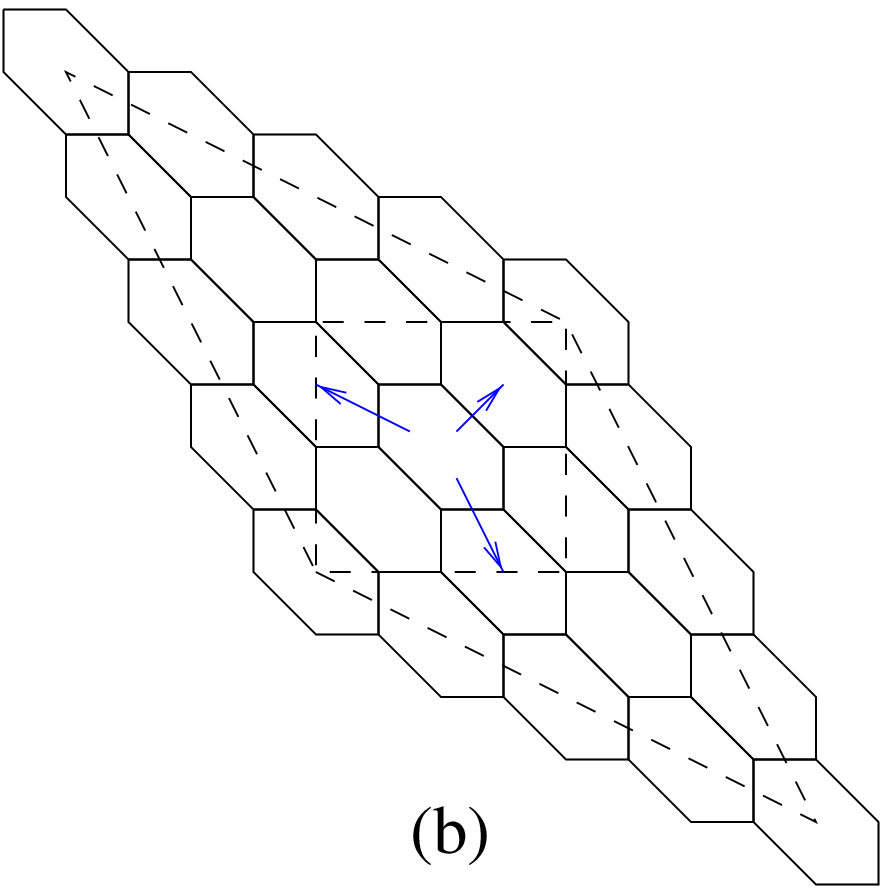}
\end{center}
\end{minipage}
\caption{Two examples of brane configurations
with a ${\bf Z}_3$ symmetry and the correspondence to the quiver diagram
of ${\bf Z}_n \times {\bf Z}_n$.
Arrows represent open strings connecting D-branes on neighboring boxes.
The small parallelogram bounded by dashed lines represents
fundamental region under the action of ${\bf Z}_3$.}
\label{fig:brane}
\end{center}
\end{figure}

We regard the correspondence
between the quiver diagram and the brane configuration
as a guideline in constructing brane configurations.
As discussed in Section 2, the quiver diagram of
$\Delta(3n^2)$ is obtained from that of
${\bf Z}_n \times {\bf Z}_n$ by the ${\bf Z}_3$ quotient.
Thus a naive guess leads to the idea that the brane configuration for
the orbifold ${\bf C}^3/\Delta(3n^2)$ is obtained from
the brane box configuration by the ${\bf Z}_3$ quotient.
However, we can not define a ${\bf Z}_3$ quotient
on the brane box configuration
since ${\bf Z}_3$ is not a symmetry of the configuration
of Figure \ref{fig:branebox}.
Instead, we construct a brane configuration for
${\bf C}^3/{\bf Z}_n \times {\bf Z}_n$
which has the ${\bf Z}_3$ symmetry
maintaining the correspondence of Table \ref{table:correspondence}.
Global structure is determined by the requirement
that it provides appropriate gauge groups
and matter contents given in Table \ref{table:correspondence}.
The requirement of the ${\bf Z}_3$ symmetry
leads to a condition on the local structure.
Naively it leads to a prong which consists of three branes
with directions $(m_1,m_2)$, $(-m_2,m_1-m_2)$ and $(m_2-m_1,-m_1)$
on a plane.
However, the ${\bf Z}_3$ equivalence relation (\ref{eq:relation})
is defined on the lattice of irreducible representations.
It means that the ${\bf Z}_3$ action on the brane configuration
is defined in reference to the "lattice of boxes".
Thus it is necessary to take directions of alignment of boxes
into account.
There are infinitely many possibilities which 
satisfy these criteria.
Two examples are depicted in Figure \ref{fig:brane}(a)
and (b).

These figures represent brane configurations on the 56-plane.
The lines indicate $(p,q)$ 5-branes which extend
along 01234 directions and
$p+\tau q$ direction in the 56-plane. Here we take $\tau=i$.
The difference between Figure \ref{fig:brane}(a) and Figure \ref{fig:brane}(b)
is charges $(p,q)$ of 5-branes.
Three types of 5-branes in Figure \ref{fig:brane}(a) have $(p,q)$ charges
$(2,1)$, $(-1,-1)$ and $(-1,-2)$,
while three types of 5-branes in Figure \ref{fig:brane}(b) have $(p,q)$ charges
$(1,1)$, $(-1,0)$ and $(0,-1)$.
Although we do not have a way to judge which is the right one at present,
we will discuss in Section 4
that the Figure \ref{fig:brane}(b) is the right one
for the brane configuraiton for ${\bf C}^3/{\bf Z}_n \times {\bf Z}_n$.
Each hexagon plays the role of each square of the brane box model.
In contrast to the brane box model, however,
D-branes on each box must be D3-branes stretching along 0156 directions
to maintain 1/8 supersymmetry.
Since D3-branes are bounded by $(p,q)$ 5-branes in 56-directions,
their effective action is two-dimensional.
This is why we started with D1-branes on orbifolds.

To realize the correspondence given in Table \ref{table:correspondence}
for $\Gamma=\Delta(3n^2)$, we must identify
the first row and the $n$-th row and the first column
and the $n$-th column.
As in the brane box model, each box gives a gauge group
$U(N_a)$ where $N_a$ is the number of D-branes on the $a$-th box.
We let the number of D1-branes on ${\bf C}^3/{\bf Z}_n \times {\bf Z}_n$
be one,
then $N_a$ becomes one for every box.
Matters come from open strings which
connect D-branes on neighboring boxes.
The presence of $(p,q)$ 5-branes induces a particular
orientation for the strings.
Relative orientations of strings are restricted by the requirement of
the invariance under the ${\bf Z}_3$ action.
There are two possibilities; one is shown in Figure \ref{fig:brane}.
The arrows indicate the orientations of strings stretching from one box.
Again only one set of three open strings is drawn
although such a set of strings stretch from every box.
These open strings reproduce the matter contents
determined by the quiver diagram
in Figure \ref{fig:quiver1} for the case
$R_3^{(m_1,m_2)}=R_3^{(1,1)}$.
Another set of orientations is obtained
by reversing the orientations of all arrows,
which is essentially equivalent to the case
shown in Figure \ref{fig:brane}.
Although we discussed orientations of open strings by the requirement
of the ${\bf Z}_3$ invariance,
it is necessary to find out a reason
that only such orientations are allowed based on string theory.

\begin{figure}[htdp]
\begin{center}
\begin{minipage}{50mm}
\begin{center}
\leavevmode
\epsfxsize=50mm
\epsfbox{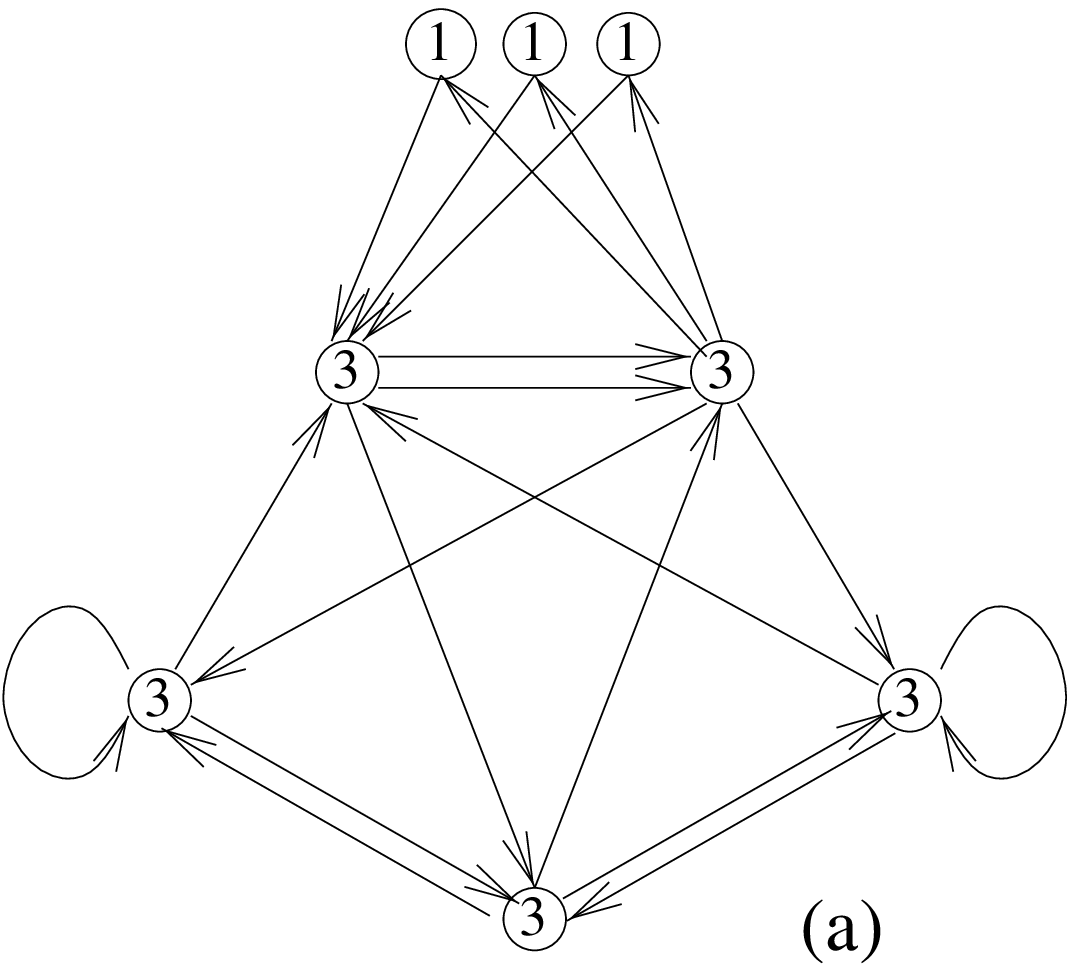}
\end{center}
\end{minipage}
\hspace{20mm}
\begin{minipage}{60mm}
\begin{center}
\leavevmode
\epsfxsize=60mm
\epsfbox{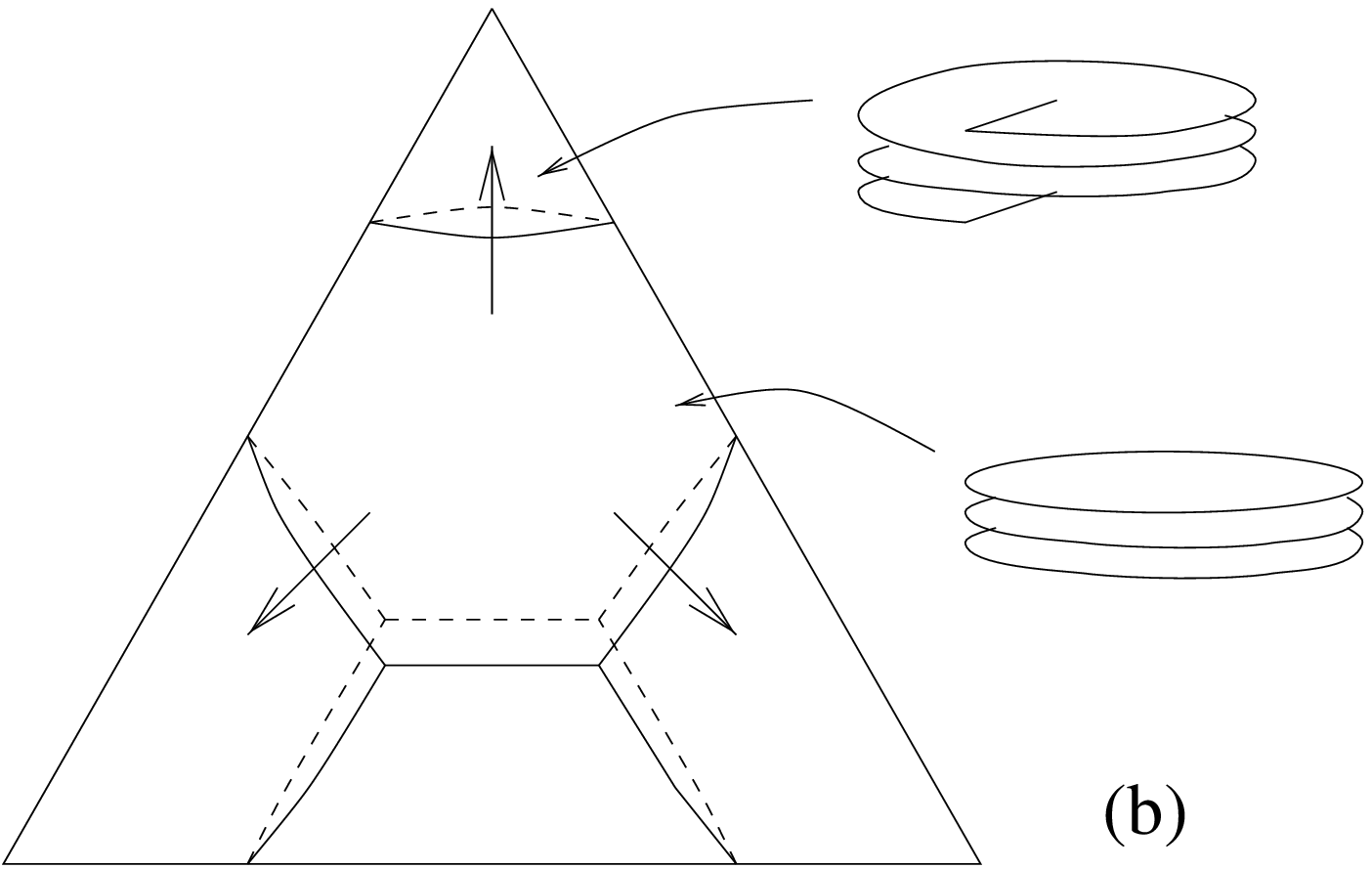}
\end{center}
\end{minipage}
\caption{(a) The quiver diagram of $\Delta(48)$.
(b) The brane configuration for ${\bf C}^3/\Delta(48)$.
Arrows represent open strings connecting D-branes on neighboring boxes.}
\label{fig:quiver-brane}
\end{center}
\end{figure}

Now that we have a brane configuration
for ${\bf C}^3/{\bf Z}_n \times {\bf Z}_n$
with a ${\bf Z}_3$ symmetry,
we make a ${\bf Z}_3$ quotient to obtain a brane configuration for
${\bf C}^3/\Delta(3n^2)$.
The ${\bf Z}_3$ quotient on the brane configuration is defined
through that on the quiver diagram.
By the ${\bf Z}_3$ quotient,
fundamental region becomes the small parallelogram
bounded by dashed lines in Figure \ref{fig:brane}.
We would like to show that the brane configuration
precisely reproduces the structure of the quiver diagram
of $\Delta(3n^2)$.
We take $n=4$ as an example.
The quiver diagram of $\Delta(3n^2)$ with $n=4$
can be depicted in a closed form as in Figure \ref{fig:quiver-brane}(a).
The brane configuration for ${\bf C}^3/\Delta(3n^2)$
can also be drawn in a closed form as in
Figure \ref{fig:quiver-brane}(b). 
It is basically a ${\bf Z}_3$ orbifold of a two-torus $T^2$. 
Three apexes in Figure \ref{fig:quiver-brane}(b)
corresponds to the fixed points of ${\bf Z}_3$.
There are three D-branes on
five boxes which do not include the fixed points,
and hence each box gives a gauge group $U(3)$.
On the other hand,
the D-brane on the box which include the fixed point
has spiral structure as illustrated in Figure \ref{fig:quiver-brane}(b)
due to the orbifolding procedure.
In this case, an oriented open string stretching from the $i$-th D-brane to
the $j$-th D-brane is the same as an open string stretching from
the $(i+1)$-th D-brane to the $(j+1)$-th D-brane (modulo 3).
Since the $(i,j)$ component of the gauge field comes from the
oriented open string from the $i$-th D-brane to the $j$-th D-brane,
the $3 \times 3$ components of the gauge field
take the following form,
\begin{equation}
A \sim \left(
\begin{array}{ccc}
a_1&a_2&a_3 \\
a_3&a_1&a_2 \\
a_2&a_3&a_1
\end{array}
\right).
\end{equation}
It means that the box including the fixed point
gives the gauge group $U(1)^3$ instead of $U(3)$.
As for the matters, they come from open strings which
connect D-branes on neighboring boxes.
Locally, every boundary between two boxes has the same structure;
there are three D-branes on both sides,
so each boundary gives a matter with $3 \times 3$ components.
The $(i,j)$ component comes from an oriented open string
stretching from the $i$-th D-brane on one side
to the $j$-th D-brane on the other side.
Difference comes from global structure of D-branes
meeting at each boundary.
It determines how the matters transform under the gauge groups.
If boxes on both side do not include fixed points of ${\bf Z}_3$,
the matter transforms as a bifundamental $3 \times {\bar 3}$
of $U(3) \times U(3)$.
If a box on one side of the boundary include a fixed point,
the corresponding representation $3$ splits into $1 \oplus 1 \oplus 1$
\footnote{If we consider $N$ D-branes on an orbifold,
the gauge group coming from the box including a fixed point
is $U(N)^3$ instead of $U(3N)$.
Accordingly, $3N$ dimensional representation splits into $N \oplus N \oplus N$;
each $N$ represents $N$-dimensional representation of each $U(N)$.}.
Thus the $3 \times 3$ components of the matter coming from such a boundary
split into three $3 \times {\bar 1}$ or three $1 \times {\bar 3}$.
They coincide with the spectrum indicated by the arrows in the quiver diagram.
Collecting these results,
one can see that the brane configuration in Figure \ref{fig:quiver-brane}(b)
presicely reproduces the structure of the quiver diagram
in Figure \ref{fig:quiver-brane}(a).

In the $n \notin 3{\bf Z}$ case,
only one of the three apexes is included in a box
and the other two lie at boundaries of some boxes
as shown in Figure \ref{fig:quiver-brane}(b).
For $n \in 3{\bf Z}$, on the other hand,
we can see that three apexes lie inside of three boxes respectively.
Therefore each box at the apex gives a gauge group factor $U(1)^3$
in accordance with the structure of the quiver diagram.

Now we would like to comment on the correspondence
between the brane configuration and the quiver diagram.
In the definition of the ${\bf Z}_3$ action on the quiver diagram,
a rather elaborate rule was necessary
in relation to the fixed points of ${\bf Z}_3$.
In contrast, 
the rule of the ${\bf Z}_3$ quotient are automatically reproduced
via the brane configurations although
the quotienting procedure on the brane configurations is simple.
In this respect, we can say that the brane configuration
inherently posesses necessary property which the quiver diagrams should have.
In other words, it seems reasonable to regard
the brane configuration as a realization of the quiver diagram
in the language of physics.

The brane configuration for
${\bf C}^3/\Delta(6n^2)$ is obtained by making a
${\bf Z}_2$ quotient in addition to the ${\bf Z}_3$
quotient for ${\bf C}^3/\Delta(3n^2)$.
The ${\bf Z}_2$ quotient is determined through
the relation (\ref{eq:relation6-2})
which defines the ${\bf Z}_2$ action on the quiver diagram.
One can see that the brane configuration reproduces
the structure of the quiver diagram of ${\bf C}^3/\Delta(6n^2)$
by a similar argument to the ${\bf C}^3/\Delta(3n^2)$ cases.

\section{Physical interpretation of McKay correspondence}
\reseteqnum

In this section, we would like to discuss relations
between the brane configurations obtained in the last section
and geometries of ${\bf C}^3/\Gamma$.
As we mentioned,
the brane configuration for ${\bf C}^3/\Gamma$
plays the role of the quiver diagram of $\Gamma$,
hence the relation can be considered as a relation
between the tensor product structure of irreducible
representations of $\Gamma$
and the geometry of ${\bf C}^3/\Gamma$.
Such a relation has been discussed in mathematics
as the McKay correspondence.
We will present evidence that the McKay correspondence may be
understood as T-duality.

Before we come to the discussion on duality,
we make it clear what we mean by the term "McKay correspondence".
The McKay correspondence was originally observed for two-dimensional
orbifolds\cite{McKay}.
It is a relation between the quiver diagram of $\Gamma$ and
a diagram representing intersections among exceptional divisors of
$\widetilde{{\bf C}^2/\Gamma}$.
Both of them coincide with a certain Dynkin diagram of $ADE$ Lie algebra.
We consider its straightforward generalization as a three-dimensional
McKay correspondence.
That is, we use the term McKay correspondence to refer to
a relation between the quiver diagram of $\Gamma$ and
a diagram which represents intersections among exceptional divisors of
$\widetilde{{\bf C}^3/\Gamma}$.
If $\Gamma$ is abelian, ${\bf C}^3/\Gamma$ becomes a toric variety
and its resolution is discussed by using toric method.
In this case, information on intersections among exceptional divisors
is represented by a toric diagram.
In summary, we use the term McKay correspondence in dimension three
as a relation between the quiver diagram of $\Gamma$ and the toric
diagram of $\widetilde{{\bf C}^3/\Gamma}$.

Fortunately,
relations between brane configurations and toric varieties
were generally discussed in \cite{LV}.
To explain the idea of \cite{LV},
we briefly review toric virieties.
A complex $d$-dimensional toric variety involves real $d$-dimensional
torus $T^d$.
Certain cycles of the torus $T^d$ shrink on some loci of the toric variety.
A toric variety is determined by specifying which cycles shrink where.
Such information is represented by a diagram in real $d$-dimensional
space, which we call a toric diagram.
Information on (co)homology of the toric variety is encoded
in the toric diagram.
For toric varieties with vanishing first Chern class,
toric diagrams are reduced to diagrams in ($d-1$)-dimensional space.
As we are considering complex three-dimensional toric varieties
with vanishing first Chern class,
their toric diagrams represent how a torus $T^2$
of the toric variety shrinks.
More precisely, what is identified with loci where $T^2$ shrinks is
a dual diagram of the toric diagram.
A line in the dual diagram with a slope $q/p$ represents
locus at which a $(p,q)$ cycle of $T^2$ shrinks.
Keeping the property of toric varieties in mind,
we turn to the explanation of \cite{LV} which relates
brane configurations and toric diagrams.
It is known that M-theory on $T^2$
and type IIB string theory on a circle $S^1$
are related by a kind of T-duality \cite{Schwarz2}.
Vanishing $(p,q)$ cycles of $T^2$ of M-theory corresponds to
$(p,q)$ 5-branes of type IIB string theory.
If we consider M-theory on a toric variety
and dualize in this sense along the 2-torus $T^2$
of the toric variety,
we obtain type IIB string theory on a circle
with  $(p,q)$ 5-branes; the skeleton of the dual diagram
is identified with a web of $(p,q)$ 5-branes.

\begin{figure}[htdp]
\begin{center}
\begin{minipage}{60mm}
\begin{center}
\leavevmode
\epsfxsize=40mm
\epsfbox{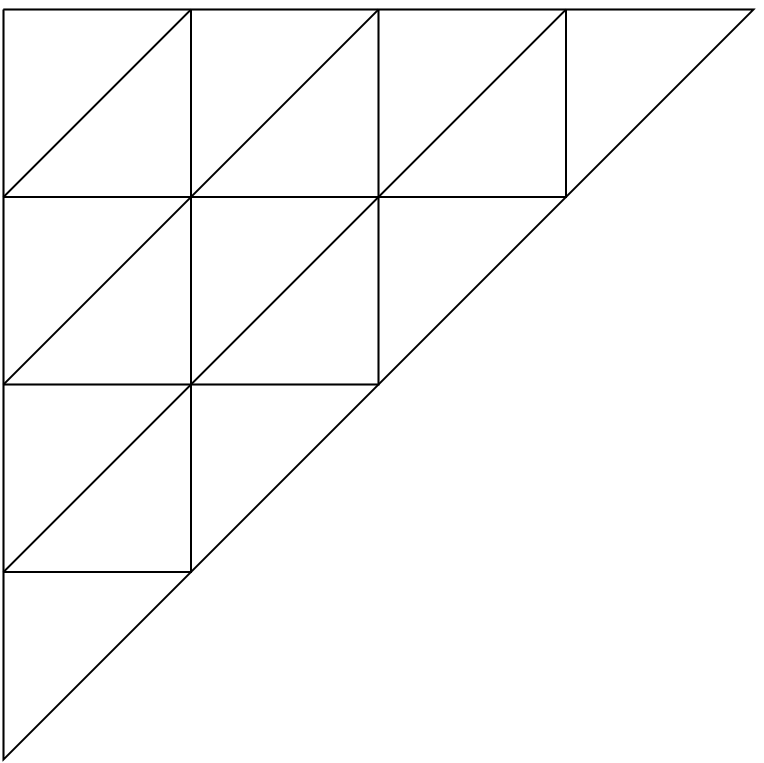}
\caption{The toric diagram for a resolution
of ${\bf C}^3/{\bf Z}_n \times {\bf Z}_n$ with $n=4$.}
\label{fig:toric1}
\end{center}
\end{minipage}
\hspace{20mm}
\begin{minipage}{60mm}
\begin{center}
\leavevmode
\epsfxsize=40mm
\epsfbox{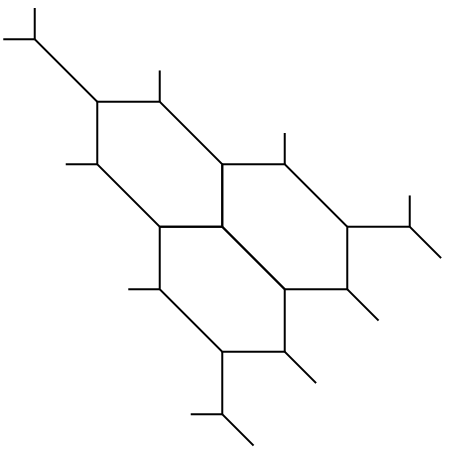}
\caption{The dual diagram of Figure \ref{fig:toric1}.}
\label{fig:dual1}
\end{center}
\end{minipage}
\end{center}
\end{figure}

We would like to apply this argument
to the case $\Gamma={\bf Z}_n \times {\bf Z}_n$.
The toric diagram of a certain resolution of the
orbifold ${\bf C}^3/{\bf Z}_n \times {\bf Z}_n$
is depicted in Figure \ref{fig:toric1}.
Its dual diagram depicted in Figure \ref{fig:dual1}
represents a web of $(p,q)$ 5-branes
in the sense described above.
The brane configuration in Figure \ref{fig:dual1} has the same
structure as the configuration for ${\bf C}^3/{\bf Z}_n \times {\bf Z}_n$
in Figure \ref{fig:brane}(b) at least locally.
In this sense, 
the brane configuration for ${\bf C}^3/{\bf Z}_n \times {\bf Z}_n$,
which can be interpreted as the quiver diagram of ${\bf Z}_n \times {\bf Z}_n$,
and the toric diagram of ${\bf C}^3/{\bf Z}_n \times {\bf Z}_n$
are related by T-duality.
The global structure of Figure \ref{fig:dual1}, however,
is different from that of Figure \ref{fig:brane}(b);
to obtain the brane configuration in Figure \ref{fig:brane}(b),
we must combine two brane configurations in Figure \ref{fig:dual1}
and make a certain identification.
The difference stems from the non-compactness of the orbifold.
The models discussed in Section 3 are so-called
elliptic models; they are compactified
along 56 directions to $T^2$.
On the other hand,
the toric diagram in Figure \ref{fig:toric1} represents
an open variety which does not have such compactness.
In fact, such a manifold is not appropriate to perform
T-duality since the radii of the circles performing T-duality
become infinite as the distance from the singularity
becomes infinite.
In order to make the T-duality argument precise,
we must replace the toric variety by a maniofold
which have the same local structure but have another
asymptotics, so that the radii of the circles are finite at infinity.
It is analogous to the replacement of ${\bf C}^2/{\bf Z}_n$
with a Taub-NUT space.
We hope that the correspondence between the brane configuration
and the toric diagram can be made definite
when such a replacement is properly taken into account.

Thus it needs further study to make the T-duality argument precise,
but there are some evidences to support such an argument.
First, we would like to point out a nontrivial coincidence
between the physical argument presented above and mathematical
argument on the McKay correspondence.
In resolving the singularity
of ${\bf C}^3/{\bf Z}_n \times {\bf Z}_n$,
there are many possibilities
related by topology changing process called a flop.
Among such resolutions, only one resolution represented by
the toric diagram in Figure \ref{fig:toric1}
has a relation to the brane configuration for
${\bf C}^3/{\bf Z}_n \times {\bf Z}_n$.
In the mathematical argument \cite{IN},
the McKay correspondence is proved
only for a paticular kind of resolution.
It is so-called the Hilbert scheme of ${\bf C}^3/{\bf Z}_n \times {\bf Z}_n$,
whose toric diagram is nothing but Figure \ref{fig:toric1}.
Although the formulation of the McKay correspondence in \cite{IN}
is different from the naive one we are considering,
it seems reasonable to regard the coincidence as a supporting evidence
for the T-duality interpretation of the McKay correspondence.

\begin{figure}[htdp]
\begin{center}
\leavevmode
\epsfxsize=40mm
\epsfbox{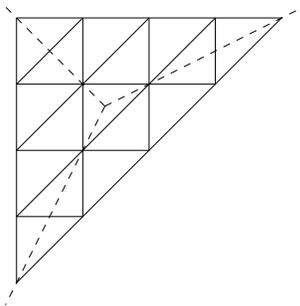}
\caption{The ${\bf Z}_3$ quotient of the toric diagram
for ${\bf C}^3/{\bf Z}_n \times {\bf Z}_n$.}
\label{fig:toric3}
\end{center}
\end{figure}

Now we turn to the case $\Gamma=\Delta(3n^2)$.
Since the orbifolds ${\bf C}^3/\Delta(3n^2)$ is not a toric variety,
we can not directly apply the above argument.
However, the brane configuration for ${\bf C}^3/\Delta(3n^2)$ is obtained
from that for ${\bf C}^3/{\bf Z}_n \times {\bf Z}_n$
by the ${\bf Z}_3$ quotient.
So we use this relation to discuss the geometry of ${\bf C}^3/\Delta(3n^2)$.
The ${\bf Z}_3$ action on the brane configuration
can be translated to the action on the toric diagram
through the relation between the Figure \ref{fig:toric1}
and Figure \ref{fig:dual1}.
For example, ${\bf Z}_3$ acts cyclically on three apexes of the triangle
which lies at the center of the toric diagram.
Since each lattice point inside the toric diagram represent
an exceptional divisor,
the ${\bf Z}_3$ quotient means that the three divisors
must be identified.
In this way, we obtain a ${\bf Z}_3$ quotient of the toric variety
${\bf C}^3/{\bf Z}_n \times {\bf Z}_n$.
The ${\bf Z}_3$ action on the quiver diagram is illustrated
in Figure \ref{fig:toric3}.

In fact, resolutions of the singularity of ${\bf C}^3/\Delta(3n^2)$
were discussed in \cite{Ito1,Ito2,Ito3}.
Broadly speaking, the singularity is resolved by the following procedure.
First, consider toric resolutions of the abelian orbifold
${\bf C}^3/{\bf Z}_n \times {\bf Z}_n$.
Then perform a certain ${\bf Z}_3$ quotient,
and finally resolve singularities which occur
due to the ${\bf Z}_3$ quotienting procedure.
The group ${\bf Z}_3$ permutes exceptional divisors
of the toric resolution of ${\bf C}^3/{\bf Z}_n \times {\bf Z}_n$.
In fact, the ${\bf Z}_3$ quotient is equivalent to
that represented in Figure \ref{fig:toric3}
if we consider Figure \ref{fig:toric1} as a resolution of
${\bf C}^3/{\bf Z}_n \times {\bf Z}_n$.
Although the ${\bf Z}_3$ actions are equivalent,
the origins are quite different.
The ${\bf Z}_3$ action in Figure \ref{fig:toric3}
were determined by the discussion on the quiver diagrams,
so it originates from the representation theory of $\Delta(3n^2)$.
On the other hand,
the ${\bf Z}_3$ action in the process of the resolution
of the singularity comes from geometrical argument.
The coincidence between the two ${\bf Z}_3$ actions
can be considered as one aspect of the McKay correspondence
for $\Gamma=\Delta(3n^2)$,
and again it serves as a supporting evidence
for the T-duality interpretation of the McKay correspondence.

\begin{figure}[htdp]
\begin{center}
\begin{minipage}{50mm}
\begin{center}
\leavevmode
\epsfxsize=45mm
\epsfbox{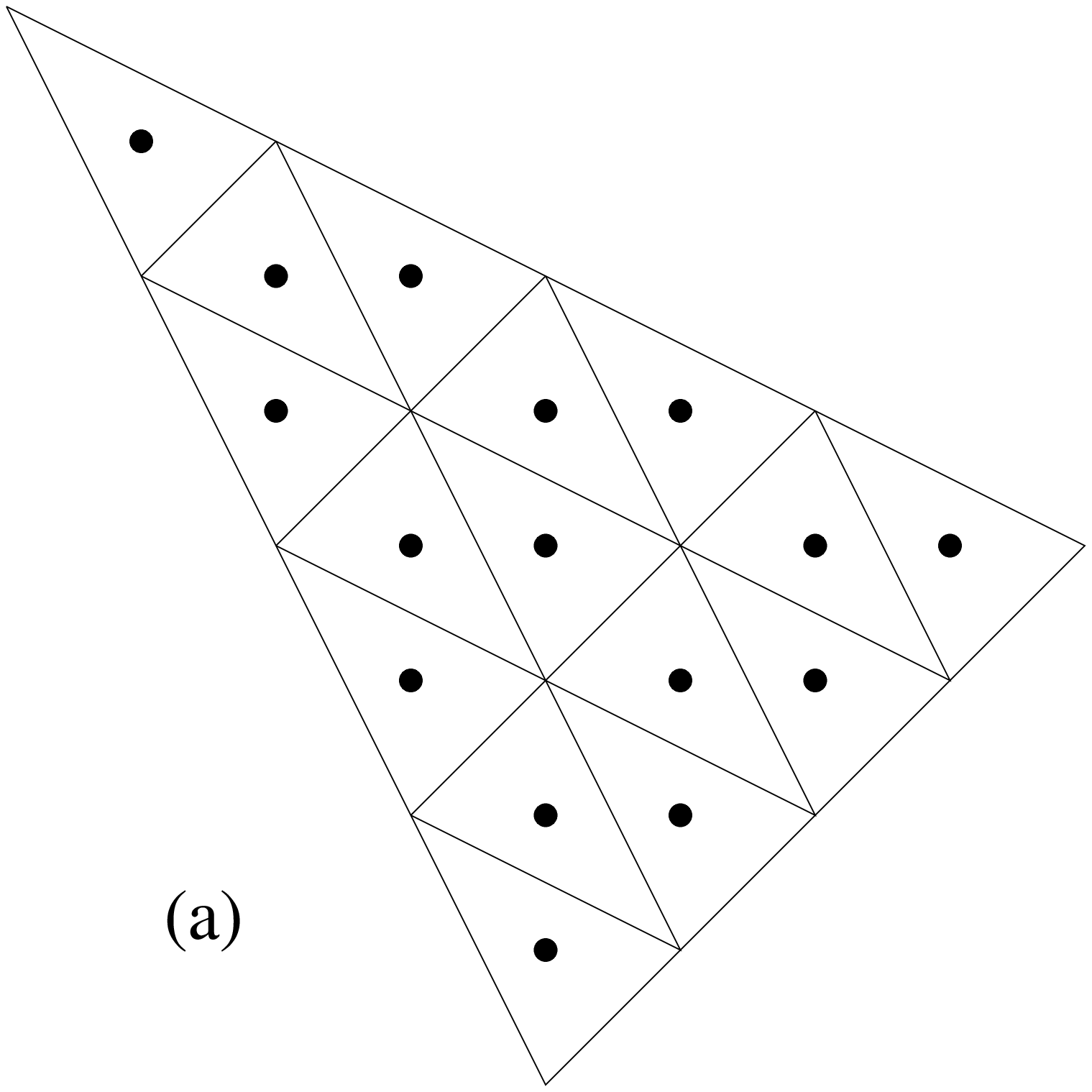}
\end{center}
\end{minipage}
\hspace{1mm}
\begin{minipage}{50mm}
\begin{center}
\leavevmode
\epsfxsize=45mm
\epsfbox{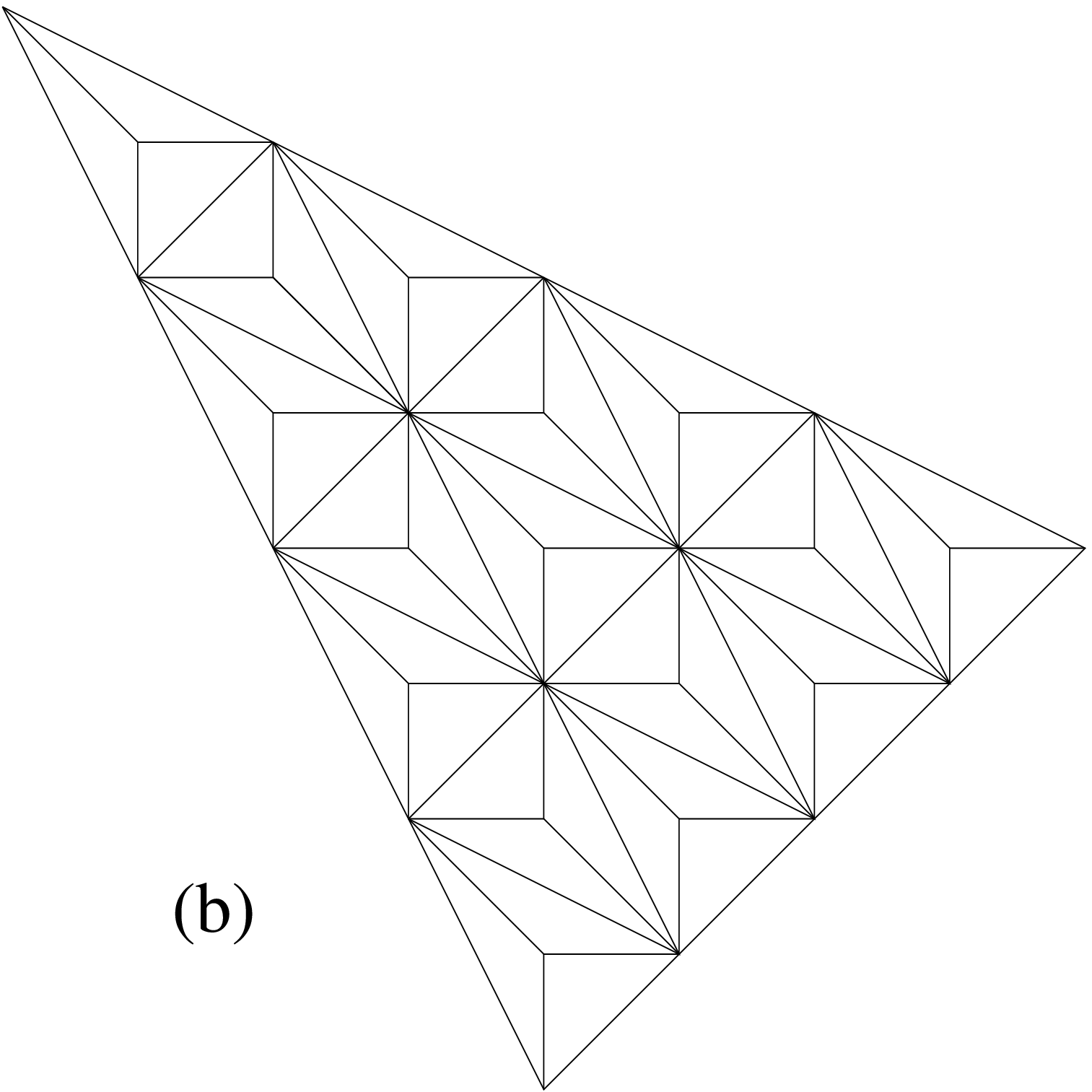}
\end{center}
\end{minipage}
\hspace{1mm}
\begin{minipage}{40mm}
\begin{center}
\leavevmode
\epsfxsize=30mm
\epsfbox{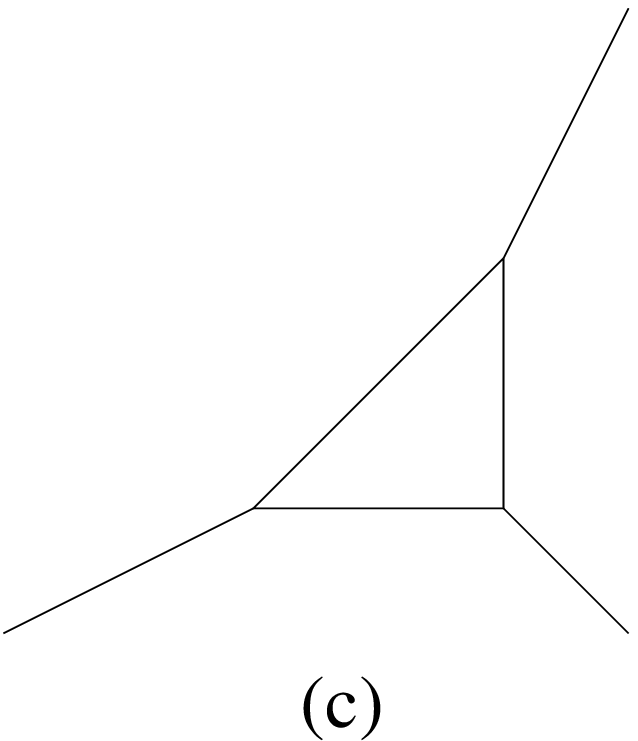}
\end{center}
\end{minipage}
\caption{(a) The toric diagram corresponding to the configuration of
Figure \ref{fig:brane}(a).
(b) The toric diagram corresponding to a space
obtained by resolving the singularity of Figure \ref{fig:resolution}(a).
(c) The local structure of a junction corresponding to
Figure \ref{fig:resolution}(b).}
\label{fig:resolution}
\end{center}
\end{figure}

Now we comment on the difference between the brane configuration of
Figure \ref{fig:brane}(a) and that of Figure \ref{fig:brane}(b).
They are only different in $(p,q)$ charges of 5-branes.
From the toric geometry point of view, however,
there is an essential difference \cite{AHK}.
The toric diagram corresponding to the configuration of
Figure \ref{fig:brane}(a) before the ${\bf Z}_3$ identification
is depicted in Figure \ref{fig:resolution}(a).
In the figure, there is a lattice point in each triangle
in contrast to Figure \ref{fig:toric1}.
It means that the orbifold singularity is not fully resolved.
To resolve fully the orbifold singularity,
we must subdivide the toric diagram as shown
in Figure \ref{fig:resolution}(b).
Then each junction of $(p,q)$ 5-branes is replaced
by Figure \ref{fig:resolution}(c)\footnote{This kind of configuration
is considered in \cite{AKLM} in the context of brane box models.}.
On the other hand, each junction in Figure \ref{fig:brane}(b)
can not be made such a replacement.
Therefore we can determine the right $(p,q)$ charges of 5-branes
by requiring that each junction can not be replaced by a set of junctions
as in Figure \ref{fig:resolution}(c).

\section{Discussions}
\reseteqnum

In this paper, we have studied brane configurations
corresponding to D-branes on nonabelian orbifolds
${\bf C}^3/\Delta(3n^2)$ and
${\bf C}^3/\Delta(6n^2)$
based on the analyses of the quiver diagrams.
We first constructed the brane configuration
for ${\bf C}^3/{\bf Z}_n \times {\bf Z}_n$
by requiring that it had a ${\bf Z}_3$ symmetry.
It consists of a web of $(p,q)$ 5-branes and D3-branes.
By making a ${\bf Z}_3$ ($S_3$) quotient,
we have obtained the brane condigurations
for ${\bf C}^3/\Delta(3n^2)$ (${\bf C}^3/\Delta(6n^2)$).
The structure of the quiver diagrams
of $\Delta(3n^2)$ and $\Delta(6n^2)$
can be naturally explained via the brane configurations.
We have also discussed relations between
the brane configurations and toric diagrams.
Based on the argument which relates branes and the toric geometry,
we have pointed out that the three-dimensional McKay correspondence
may be understood as T-duality.

In \cite{HU}, it was shown that
D3-branes on orbifolds and the brane box model are
related by T-duality.
Now we would like to make a rough argument to relate
the brane configurations and D-branes on orbifolds.
We start with type IIB string theory with a web of
$(p,q)$ 5-branes along 01234 directions and one direction of the
56 plane and D3-branes along 0156 directions.
We first T-dualize along direction 9
and decompactify along direction 10.
Then we obtain M-theory on an orbifold ${\bf C}^3/\Delta(3n^2)$
with M5-branes along 01569(10) directions.
Here ${\bf C}^3/\Delta(3n^2)$ extends to 56789(10) directions.
To obtain type IIB string theory on orbifolds,
we take a limit that the direction 1 shrinks.
Then we obtain type IIA string theory on ${\bf C}^3/\Delta(3n^2)$
with D4-branes along 0569(10).
Next we perform T-dualities along 256 directions.
Then we obtain type IIB string theory on ${\bf C}^3/\Delta(3n^2)$
with D3-branes.
Here ${\bf C}^3/\Delta(3n^2)$ extends to 56789(10) directions,
while D3-branes extend to 029(10) directions.
If such an argument on duality is true,
the right interpretation of the D1-branes
discussed in Section 2
seems to be D3-branes wrapped around two directions
of the orbifold.
D-branes wrapping on each cycle
of an orbifold stick to the singularity \cite{Douglas,DDG}.
However, since we are considering a set of D-branes which corresponds
to the regular representation of $\Delta(3n^2)$,
they are allowed to move away from the singularity \cite{HU}.
Anyway it needs further investigation on the argument of duality.

One of the important problems which should be clarified
is the structure of the Kahler moduli space of nonabelian orbifolds.
For three-dimensional abelian orbifolds,
their Kahler moduli spaces were examined in \cite{DGM}
by using a toric method
combining with the fact that toric varieties can be realized
as vacuum moduli spaces of two-dimensional
gauged linear sigma models \cite{Witten}.
On the other hand, 
this method can not be applied to nonabelian orbifolds
since they are not toric varieries.
The nonabelian orbifolds, however, are related to toric varieties
through a quotienting procedure.
It would be interesting to investigate whether
the quotienting procedure can be incorporated into the toric method
to analyze the structure of Kahler moduli
space of nonabelian orbifolds.


It is also interesting to generalize the work
to other cases.
For the  E-type subgroups of $SU(3)$,
a catalogue of quiver diagrams is given in \cite{HH1}.
It is so complicated that construction of brane configurations
seems to be difficult.
However, unbroken supersymmetry may provide us a guide
to the construction.
For $\Gamma \subset SU(4)$,
classification of finite subgroups were summarized in \cite{HH2}.
Since there is no supersymmetry in these cases,
quiver diagrams representing bosonic matter contents
and those representing fermionic matter contents are different.
It would be interesting to find a mechanism
to provide such asymmetric matter contents from brane configurations.

\vskip 1cm
\centerline{\large\bf Acknowledgements}

I would like to thank Y.Ito, S. Hosono, T. Kitao, Y. Sekino,
T. Hara and S. Sugimoto
for valuable discussions and comments.
I wish to express my special thanks to T. Tani
for helpful suggestions on several points in the paper.
This work is supported in part by Japan Society for the
Promotion of Science(No. 10-3815).

\end{document}